\documentclass[twocolumn]{aastex62}
\usepackage{lineno, blindtext}
\usepackage{epsfig}
\usepackage{natbib}
\usepackage{amsmath}
\usepackage{verbatim}
\definecolor{blue}{RGB}{50, 80, 255}

\shorttitle{Discovery of Be in White Dwarfs}
\shortauthors{B. Klein et al.}

\begin{document}

\title{Discovery of Beryllium in White Dwarfs Polluted by Planetesimal Accretion}

\correspondingauthor{Beth Klein}
\email{kleinb@astro.ucla.edu}

\author[0000-0001-5854-675X]{Beth Klein}
\affiliation{Department of Physics and Astronomy, University of California, Los Angeles, CA 90095-1562, USA}

\author[0000-0003-0053-3854]{Alexandra E. Doyle}
\affiliation{Earth, Planetary, and Space Sciences, University of California, Los Angeles, Los Angeles, CA 90095, USA}

\author[0000-0001-6809-3045]{B. Zuckerman}
\affiliation{Department of Physics and Astronomy, University of California, Los Angeles, CA 90095-1562, USA}

\author[0000-0003-4609-4500]{P. Dufour}
\affiliation{Institut de Recherche sur les Exoplan\`etes (iREx), Universit\'e de Montr\'eal, Montr\'eal, QC H3C 3J7, Canada}
\affiliation{D\'epartement de physique, Universit\'e de Montr\'eal, Montr\'eal, QC H3C 3J7, Canada}

\author[0000-0002-9632-1436]{Simon Blouin}
\affiliation{Los Alamos National Laboratory, P.O. Box 1663, Mail Stop P365, Los Alamos, NM 87545, USA}

\author[0000-0001-9834-7579]{Carl Melis}
\affiliation{Center for Astrophysics and Space Sciences, University of California, San Diego, CA 92093-0424, USA}

\author[0000-0001-6654-7859]{Alycia J. Weinberger}
\affiliation{Earth and Planets Laboratory, Carnegie Institution for Science, 5241 Broad Branch Rd NW, Washington, DC 20015, USA}

\author[0000-0002-1299-0801]{Edward D. Young}
\affiliation{Earth, Planetary, and Space Sciences, University of California, Los Angeles, Los Angeles, CA 90095, USA}
 

\begin{abstract}

The element beryllium is detected for the first time in white dwarf stars. 
This discovery in the spectra of two helium-atmosphere white dwarfs was 
made possible only because of the remarkable overabundance of Be 
relative to all other elements, heavier than He, observed in these stars.
The measured Be abundances, relative to chondritic, are by far the largest 
ever seen in any astronomical object.  We anticipate that the Be in these accreted planetary bodies was 
produced by spallation of one or more of O, C, and N in a region of high
fluence of particles of MeV or greater energy.
\\
\end{abstract}

\section{Introduction \label{sec:intro}}
Over the past decade or so, the presence of absorption lines from elements heavier than helium in the spectra of single white dwarfs (WDs) with $T_{\rm eff}$ $\lesssim$25,000\,K, has come to be understood without a doubt as the result of accretion of disrupted planet(esimal)s from their ancient planetary systems \citep[e.g. reviews by][]{jurayoung2014, veras2016, farihi2016review, zuckermanyoung2018}\footnote{with the exception of the subset of WDs where carbon has been dredged-up from the interior \citep{koester1982}.}.
As a result, our knowledge and understanding of the compositions of exoplanets has grown significantly through the extraordinary detail and precision afforded by this powerful observational technique. To date, exoplanetesimal compositions measured in WD atmospheres that are `polluted' with accreted material are mostly similar to rocky bodies in the solar system, with some interesting exceptions that include water-rich bodies \citep{farihi2011gd61, farihi2013, raddi2015, xu2017, gentilefusillo2017, hoskin2020}.   Although nothing as bizarre as a carbon-dominated planet \citep[e.g.][]{bond2010} has ever been revealed in studies of WDs, significant variations in overall element-to-element abundance ratios have been measured among various WDs.  Such variations are generally attributed to sampling of material affected by igneous differentiation, i.e.~originating primarily from the crust and mantle \citep{zuckerman2011, melisdufour2017} or core \citep{melis2011, gaensicke2012cos} portions of a differentiated rocky body.   

However, what has previously not been seen is a dramatic deviation of the abundance of a specific element from an (understandable) overall pattern of elements.   In the present paper we report the discovery of beryllium (Be) with remarkably high abundances relative to those of the elements magnesium, silicon, and iron in two WDs:  GALEX 2667197548689621056 (23:39:17.03, $-$04:24:24.7, J2000; hereafter GALEXJ2339) and GD 378 (18:23:37.01, +41:04:02.1, J2000).  In both stars the Be is over-abundant by about two orders of magnitude. This is, far and away, the largest abundance of Be relative to chondritic ever measured in any astronomical object.

\begin{table*}
\caption{History of Element Discovery from Accreted Planetary Material in White Dwarfs cooler than 25,000\,K \label{tab:history}}
\begin{center}
\begin{tabular}{lllcrl}
\hline 
\hline
Discovery	&	Polluting 	&	WD	&	Facility	&	Reference	&	Notes	\\
year	&	Element(s)									\\
\hline											
1917	&	Ca	&	vMa2	&	Mt Wilson	&	vM1917	&	First ever exoplanet evidence	\\
1941	&	Mg?	&	Ross 640	&	McDonald	&	K1941	&	``probably Mg"	\\
1956	&	Fe	&	vMa2	&	Hale	&	G1956	&	blended Fe {\small I}	\\
1960	&	Mg	&	vMa2	&	Hale	&	W1960	&	Fig.~6 W1960	\\
1976	&	Na	&	G165-7	&	Hale	&	G1976	&	Fig.~1 G1976	\\
1980	&	Si	&	Ross 640	&	IUE	&	CG1980	&		\\
1980	&	Cr	&	G165-7	&	Lick/IDS	&	WL1980	&		\\
1991	&	C?	&	G238-44	&	IUE	&	V1991	&	Sec 2.1 H1997	\\
1995	&	C, O?, Al?	&	GD 40	&	HST/FOS	&	S1995	&	O \& Al unclear	\\
1998	&	Al	&	G238-44	&	IUE	&	H1998	&		\\
2007	&	Sc, Ti, V, Mn, 	&	GD 362	&	Keck/HIRES	&	Z2007	&	Earth/Moon-like	\\
	&	Co, Ni, Cu, Sr	&		&		&		&	composition	\\
2008	&	O, S	&	GD 378, GD 61	&	FUSE	&	D2008	&	unambiguous O	\\
2012	&	P	&	GD 40, G241-6, GALEXJ1931	&	HST/COS	&	J2012, G2012	&		\\
2017	&	N	&	G200-39	&	HST/COS	&	X2017	&	extrasolar KBO	\\
2020	&	Li, K	&	WDJ1644 (+others w/Li)	&	SOAR/Goodman	&	K2020	&		\\
2021	&	Li, K	&	LHS 2534 (+others w/Li)	&	VLT/X-shooter	&	H2021	&		\\
2021	&	Be	&	GALEXJ2339, GD 378	&	Keck/HIRES	&	this paper	&	spallation	\\
\hline	
\end{tabular}
\end{center}
\tablecomments{We trace the discovery of elements that unambiguously come from accretion of planetary material.  To that end we restrict this discovery timeline to WDs cooler than 25,000\,K so as to avoid confusion with other processes such as radiative levitation (see discussion in text). The discovery year is associated with the paper in which unambiguous spectral features associated with a given element were first identified.  In many cases an abundance analysis came later. The comment ``(+ others w/Li)'' in the two `Li, K' rows indicates that there are additional WDs in those studies in which Li (but not K) were detected. References are: vM1917 \citep{vanMaanen1917}; K1941 \citep{kuiper1941}; G1956 \citep{greenstein1956}; W1960 \citep{weidemann1960}; G1976 \citep{greenstein1976}; CG1980 \citep{cottrellgreenstein1980}; WL1980 \citep{wehrseliebert1980}; V1991 \citep{vts1991}; S1995 \citep{shipman1995}; H1997 \citep{holberg1997}; H1998 \citep{holberg1998}; Z2007 \citep{zuckerman2007}; D2008 \citep{desharnais2008}; J2012 \citep{jura2012cos}; G2012 \citep{gaensicke2012cos}; X2017 \citep{xu2017}; K2020 \citep{kaiser2020}; H2021 \citep{hollands2021}. 
}
\end{table*}

Since the earliest observations of WDs a century ago, a growing list of accreted elements has been fundamental in understanding WD pollution and associated cosmochemical insights.
Similar to many fields in Astronomy, the progress over time for new detections generally traces the capabilities enabled by the available observational facilities, as well as the commitment by observers to consider the best targets.   Table \ref{tab:history} is an attempt to summarize the progress of new element identifications in white dwarf atmospheres polluted by accreted planetary material.

The first identifications of white dwarf pollution were found in optical spectra through the Ca {\small II} H and K resonance lines in cool helium-atmosphere WDs\footnote{Ca equivalent widths in these stars can be $\gtrsim$ 40\,{\AA}   \citep[e.g.][]{liebert1987}.} beginning with the iconic vMa2 \citep{vanMaanen1917}. While van Maanen was looking for companions to high-proper-motion stars, he unwittingly observed a polluted white dwarf.  It took much longer (almost a century!) until it was appreciated that van Maanen's observation was in fact the first evidence of the existence of an extrasolar planetary system \citep{zuckerman-vMa2, farihi2016review}. 

Mg and Fe were next to be identified in optical spectra \citep{kuiper1941, greenstein1956}, followed later by Na \citep{greenstein1976}. Then the International Ultraviolet Explorer (IUE) satellite opened up access to UV wavelengths where elements such as Si could be detected \citep{cottrellgreenstein1980}.  Also, since Mg and Fe are more readily detected in this spectral region, the IUE significantly increased the number of known polluted WDs of the time \citep[review by][]{koester1987}.  Meanwhile, the minor element Cr was identified in optical spectra by \citet{wehrseliebert1980}. 
 
As early as 1981, many of the major and minor elements, such as:  C, N, O, Al, Si, P, S, Mn and Ni were found in ultraviolet spectra of very hot WDs (effective temperature $T_{\rm eff}$ $\gtrsim$ 60,000\,K) \citep{bruhweilerkondo1981, dupreeraymond1982, sion1985,  holberg1993, holberg1994, vennes1996}. However, since radiative levitation can have an important effect at effective temperatures higher than 25,000\,K \citep[][and references therein]{koester2014da}, the origin of heavy elements in such stars is unclear.  We are aware of analyses which argue that heavy elements in some hot WDs are accreted \citep[e.g.][]{vennes1996, barstow2014, wilson2019gd394, schreiber2019}, although from what source(s) remains uncertain.

Referring to WDs cooler than 25,000\,K, carbon was observed in IUE spectra of some WDs with $T_{\rm eff}$ $<$ 13,000K \citep[e.g.][]{wegner1981}, but those turned out to be due to dredge-up \citep{koester1982}, not planetesimal accretion.  As far as we know, in WDs cooler than 25,000\,K, carbon accreted from planetary material was first unambiguously identified in GD 40 via Hubble Space Telescope (HST) Faint Object Spectrograph (FOS)  observations by \citet{shipman1995}.  \citet{shipman1995} also reported possible, but uncertain/blended, detections of O and Al.
Al was clearly detected in G238-44, by IUE, as reported in the survey by \citet{holberg1998}, while the Far Ultraviolet Spectroscopic Explorer (FUSE) satellite brought the unambiguous identifications of O and S in GD 378 and O in GD 61 \citep{desharnais2008}\footnote{In re-examination, lines of P are also present in the FUSE spectrum of GD 378, but were not identified by \citet{desharnais2008}.}.  

A dramatic breakthrough occurred when the High Resolution Echelle Spectrograph \citep[HIRES,][]{vogt1994} on the Keck 1 Telescope was used to observe the extremely polluted WD, GD 362. Those spectra displayed absorption lines of 15 elements heavier than He \citep{zuckerman2007} including the trace elements Sc and Sr with abundances nine orders of magnitude less than H.  The pattern of element abundances in this WD led to the conclusion that it had accreted a planetary body with composition similar to a rocky planetesimal. 
HST/COS added P in GD 40, G241-6, and GALEXJ1931+0117 \citep{jura2012cos, gaensicke2012cos}, and N in G200-39 \citep[=WD1425+540,][]{xu2017}, where in the latter case, the polluting parent body is an extrasolar Kuiper Belt analog.  Note that G200-39 has a common proper motion main sequence companion, G200-40, and sometimes their names have been confused in the literature.

\begin{figure}
\includegraphics[width=80mm]{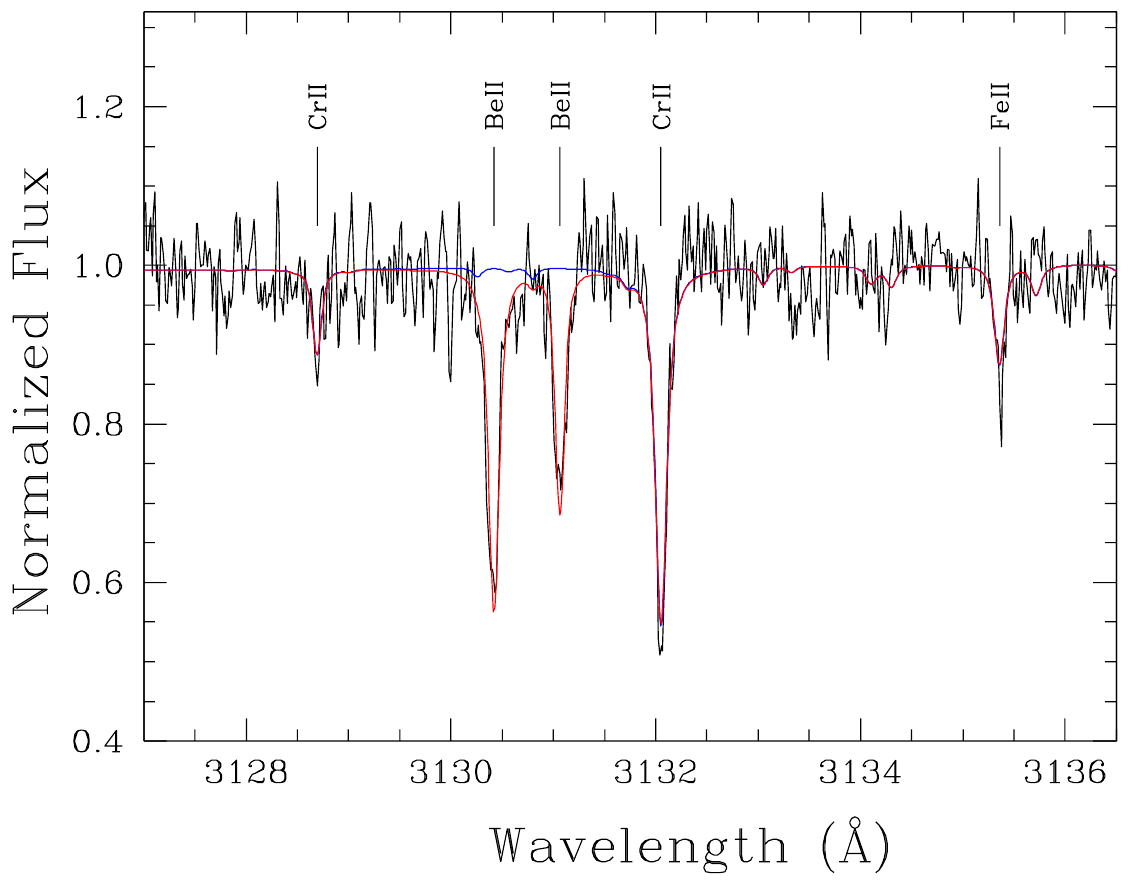}
\caption{Be {\small II} detection in GALEXJ2339 from Keck/HIRES. Wavelengths are in air and the laboratory rest frame. The data are plotted in black, and the red line is our best-fit model.  The blue line is the same model, but with the abundance of Be set to zero, demonstrating that the absorption features at 3130.42 and 3131.06\,{\AA} come from Be, without significant contribution from other elements.}
\label{fig:Be_2339}
\end{figure}

\begin{figure}
\includegraphics[width=80mm]{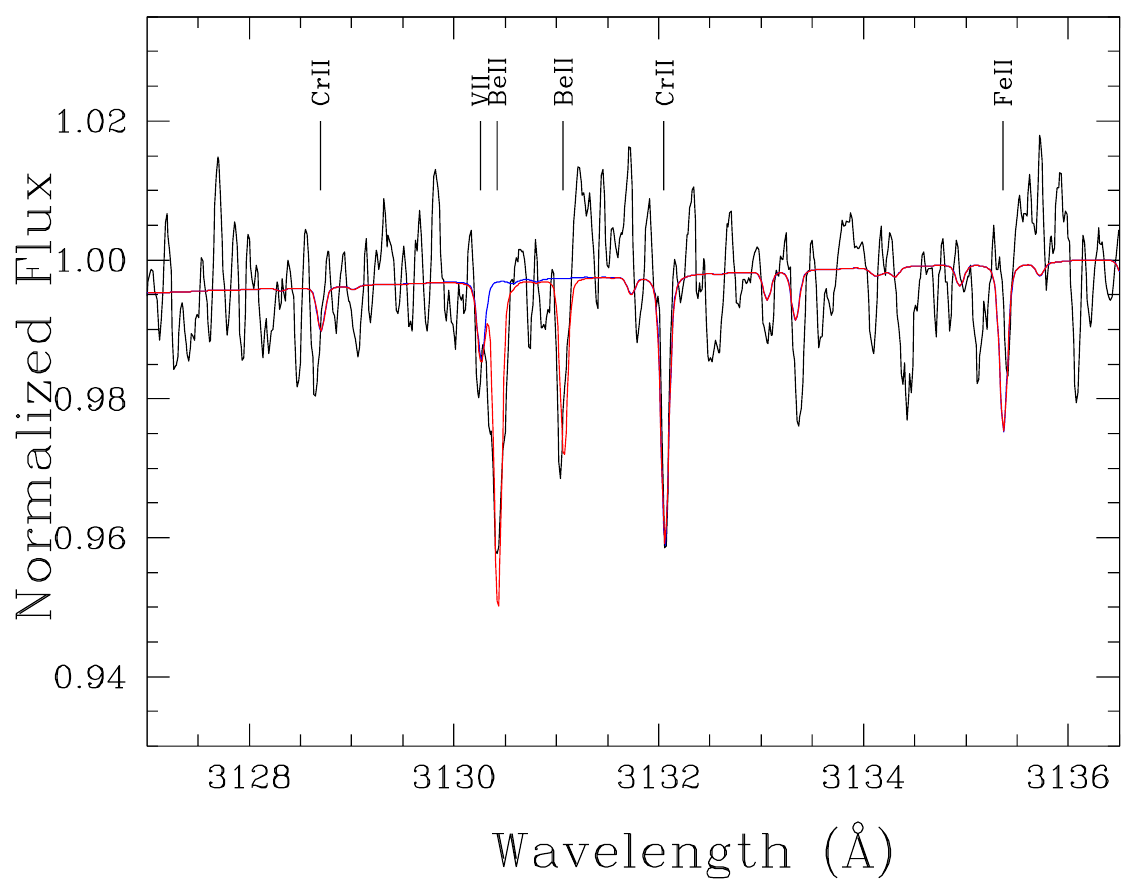}
\caption{Be {\small II} detection in GD 378, similar to Figure \ref{fig:Be_2339}, but smoothed by a 5-point boxcar average for clarity.  A possible small contribution from V {\small II} 3130.26\,{\AA} to the blue wing of the Be {\small II} 3130.42\,{\AA} line is shown in our model, but this is only based on a measured upper limit for V.}
\label{fig:Be_gd378}
\end{figure}

Recently, with parallax and photometry measurements from Gaia DR2 \citep{gaiacollaboration2016, gaiacollaboration2018}, it has been possible to identify previously unknown WDs from photometric colors and absolute magnitudes.  Thanks to those data, studies covering the cooler end of the WD sequence have recently resulted in the detections of Li and K, by two independent groups \citep{kaiser2020, hollands2021}. In this paper, using Gaia DR2 along with the exquisite sensitivity of Keck/HIRES at wavelengths as short as 3130\,{\AA}, we have the first detection of Be in the atmospheres of two polluted WDs: GALEXJ2339 and GD 378 (Figures \ref{fig:Be_2339} and \ref{fig:Be_gd378}).

GD 378 has been known for a long time as a WD with a helium-line dominated spectrum (spectral type\footnote{Spectral classifications in this paper follow the system of \citet{sion1983}: the first letter in a white dwarf type is D (degenerate star), followed by letters representing the optically detected presence of spectral features from:  hydrogen (A),  helium (B),  elements heavier than H or He (Z) in decreasing order of the observed line strengths.} DB), identified half a century ago  by \citet{greenstein1969} and later found to also display hydrogen lines (DBA) \citet{greenstein1984}. The detection of calcium absorption in GD 378 elevated it to the category of being one of the first of three known DBAZ stars, including GD 61 and G200-39 \citep{sion1988, kenyon1988}.

In contrast to the well-known GD 378, the newly found GALEXJ2339 was only recently identified as a WD \citep[coincidently in time, by the spectroscopic efforts described in this paper and the 100 pc sample of][]{kilic2020}, thanks to Gaia DR2 and the subsequent assembly of WD-candidate catalogs \citep[e.g.][]{gentilefusillo2019, jimenez-esteban2018}. Low resolution optical spectroscopy of GALEXJ2339 reveals the presence of H, He, and elements of higher Z.    Since the H lines are the strongest optical features, followed by the He lines, and then the  Ca {\small II} K-line, we classify this star as spectral type DABZ.  Nonetheless, our atmosphere models show that the composition is predominantly helium.  Thus GALEXJ2339 is another example of a heavily polluted WD with a helium-dominated atmosphere, but whose spectral type starts with `DA', as the Balmer hydrogen lines are the strongest features in its optical spectrum\footnote{There appears to be some confusion in the literature recently in which classical DZ stars such as Ross 640, L745-46A, and PG1225$-$079 have been referred to as DA white dwarfs and/or categorized as having the Balmer lines `dominate' their spectral classification. Our observation is that these particular WDs are not dominated optically by hydrogen lines since the Ca {\small II} H\&K lines are the strongest features in the optical spectra. Following the WD classification system of \citet{sion1983} they are DZA(B).}, with equivalent widths (EWs) greater than those of either the He or heavy element (Ca {\small II} H \& K) lines \citep[e.g.][]{koester2005gd16, zuckerman2007, raddi2015}.

 High resolution optical spectroscopy of GALEXJ2339 reveals absorption lines of many elements, with the surprising appearance of two relatively strong lines from beryllium (Figure \ref{fig:Be_2339}), {\it an element that has not been seen before in a WD of any type}.  

The visibility of the Be lines in GALEXJ2339 reminded us of a previously noted hint of a line in our 2008 HIRES spectrum of GD 378, near the wavelength of the strongest Be line.  This led to a follow-up HIRES observation which improved the signal-to-noise (SNR) enough to detect both Be lines in GD 378 (Figure \ref{fig:Be_gd378}).  We also re-examined datasets of other heavily polluted WDs to check for any similar features which may have been overlooked, but we found no other obvious detections of Be.  Our abundance analyses for GALEXJ2339 and GD 378 described here show the measured Be ratios relative to other elements to be extraordinary in both stars, approximately two orders of magnitude higher than those measured in main sequence stars and in meteorites.

This paper is organized as follows. In Section \ref{sec:obs} we list our observations that led to this discovery. Our atmosphere models are described in Section \ref{sec:model} along with plots of the detections of all elements in the spectra. Section \ref{sec:abund} provides an analysis of the calculated abundances.   In Section \ref{sec:discussion} we discuss our findings and conclusions, but we also refer readers to the companion paper to this one by \citet{doyle2021} for an extensive interpretation of our findings.  Some topics regarding model fitting and uncertainty calculations are detailed in Appendices A and B.

\section{Observations \label{sec:obs}}

\subsection{GALEXJ2339}

\subsubsection{KAST \label{sec:kast_j2339}}

We first observed GALEXJ2339 on 2018 December 28 (UT) with the KAST Spectrograph\footnote{\url{https://mthamilton.ucolick.org/techdocs/instruments/kast/}} on the 3 m Shane telescope at Lick Observatory as part of our large scale survey to search for heavily polluted WDs among newly identified WD candidates from Gaia DR2 \citep{gentilefusillo2019, melis2018}.   
Our setup employed a 2\arcsec~slit with the d57 dichroic splitting light to the blue arm 600/4310 grism and red arm 830/8460 grating, providing coverage of 3450-7800\,{\AA} and a resolving power $R$ = $\lambda \over \Delta\lambda$ $\simeq$700 in the blue and $\simeq$2300 in the red.

The data were reduced using standard IRAF \citep{iraf1986} slit spectra routines. The resulting spectrum has a SNR per pixel of $\simeq$35 near 4500\,{\AA} and $\simeq$26 near 7500\,{\AA}.  Absorption lines from both H {\small I}  and He {\small I} are clearly detected and the Ca {\small II} K-line and Mg {\small II} 4481\,{\AA} are apparent in these data.  

A second KAST spectrum was obtained on 2019 July 12, with a shifted red arm setup that covers the Ca infrared triplet ($\lambda$ 8498/8542/8662\,{\AA}) to check if GALEXJ2339 might be another addition to the small, but growing, set of gas disk emission systems \citep{melis2020, dennihy2020arXiv, gentilefusillo2020arXiv}.  A 1.5\arcsec~slit was used with a similar instrument setup as described above, except with the 830/8460 grating position shifted toward the red.  The resulting red arm spectrum covers 6440$-$8750\,{\AA} at a resolving power of $\simeq$2300 with a SNR $\simeq$42 near 7500\,{\AA}. In addition to the various H {\small I}  and He {\small I} lines, the O {\small I} 7772\,{\AA} triplet (unresolved), as well as a weak O {\small I} 8446\,{\AA} feature, are detected in this spectrum.  No emission was detected from the Ca {\small II} infrared triplet or any other parts of GALEXJ2339's spectrum.

\subsubsection{MagE \label{sec:mage_j2339}}
A moderate resolution optical spectrum of GALEXJ2339 was acquired with the MagE echellette spectrograph on the Magellan 1 (Baade) telescope at Las Campanas Observatory on 2019 July 03. The target was observed  for 3000 s  through the 0.5\arcsec~slit, for a resolving power of $R$ $\simeq$ 8000. Data reduction including flat fielding, spectral extraction, and wavelength calibration was performed with the Carnegie Python pipeline \citep{kelson2000, kelson2003}.  All in one shot, these MagE data covered wavelengths from 3110 to 10,000\,{\AA} with a SNR of $\simeq$10 at 3130\,{\AA} (near the Be lines), $\simeq$50 at 4485\,{\AA},  $\simeq$65 near 5170\,{\AA},  $\simeq$30 near 7770\,{\AA}, and $\simeq$25 near 8500\,{\AA}.  This spectrum confirmed the detection of O {\small I} 7772, the non-detection of gas-emission features from the Ca {\small II} infrared triplet, and discovered absorption lines from the Mg {\small I} 3838 triplet, Mg {\small I} 5184, Ca {\small II} 3706/3737, Si  {\small II} 3856/3863,  Si {\small II} 6347/6371, Ca {\small II} 8542/8662, and a hint of Mg {\small II} 7896.  

\subsubsection{HIRES \label{sec:hires_j2339}}
On 2019 July 07 (UT) we used the High Resolution Echelle Spectrograph \citep[HIRES,][]{vogt1994} on the Keck 1 Telescope  at Mauna Kea Observatory, configured with the blue collimator to observe GALEXJ2339.  The C5 decker (slit width 1.148\arcsec) provided a resolving power of $R$ $\simeq$ 40,000 and wavelength coverage of 3120 - 5950\,{\AA}. Data were reduced with similar methods and routines as described in \citet{klein2010}, using both IRAF and the MAKEE\footnote{\url{https://www.astro.caltech.edu/~tb/makee/}} software reduction packages.  In excellent conditions (clear skies, 0.6\arcsec~seeing), a 3000 s integration resulted in a spectrum with SNR $\simeq$23 around 3130\,{\AA} near the Be lines, $\simeq$50 near 4481\,{\AA}, and $\simeq$35 near 5170\,{\AA}.   On 2020 Oct 07 we obtained two hours of integration with the HIRES red collimator and C5 decker (again with clear skies, 0.6\arcsec~seeing), resulting in a spectrum of wavelengths from 4750 - 9000\,{\AA} with SNR $\simeq$80 near 5170\,{\AA}, $\simeq$45 near 7770\,{\AA}, and $\simeq$40 near 8500\,{\AA}.

\subsection{GD 378}

\subsubsection{HIRES}
GD 378 was observed with the HIRES blue collimator for 3900 s on 2008 Feb 13 (UT) and red collimator for 2400 s on 2008 Feb 26 (UT). Observing conditions were good with 0.7\arcsec~seeing on 2008 Feb 13 and 0.8\arcsec~on 2008 Feb 26.   The instrument setups and data reduction were similar as described above, although for GD 378's red data, an additional re-normalizing processing step was applied to calibrate and remove second order flux contamination in the region 8200 - 9000\,{\AA} as described in \citet{klein2010} and \citet{melis2010}.

Recently, on 2020 Oct 08 with clear skies and 0.6\arcsec~seeing, we obtained an additional 4000 s integration with the HIRES blue collimator. This resulted in a SNR of 90 near 3130\,{\AA} and $\simeq$150 near 5170\,{\AA} in the final co-added blue spectra.  The SNR of the red spectrum is $\simeq$85 near 5170\,{\AA}, $\simeq$50 near 7770{\AA}, and $\simeq$32 near 8500\,{\AA}. 

\subsubsection{FUSE \label{sec:fuse}}
Far ultraviolet observations of GD 378 were made with the FUSE satellite on 2004 May 22 (UT) under program ID D168 (PI: F. Wesemael).  The wavelength range is 905 - 1185\,{AA} with $R$ $\simeq$ 15,000.  These UV data provide coverage of elements (especially volatiles) that are generally not accessible from ground-based optical facilities, so we decided to re-analyze the FUSE spectrum with our updated model codes and abundance fits in coordination with our HIRES data. We downloaded the spectra from the Mikulski Archive for Space Telescopes (MAST), which were processed using the {\small CALFUSE} pipeline version 3.2.  \citet{desharnais2008} previously published an abundance analysis of the FUSE data including identifications of C, O, Si, S, and Fe.  In the present paper  we confirm the identifications of these elements, and we also identify phosphorous in the photosphere of GD 378.

\clearpage
\startlongtable
\begin{deluxetable}{lcc|lccl}
\tablecaption{Absorption Lines in GALEXJ2339 \label{tab:linelist_j2339}}
\tablehead{
\colhead{Ion} & \colhead{$\lambda$} & \colhead{EW} & \colhead{Ion} & \colhead{$\lambda$} & \colhead{EW} \\
\colhead{} & \colhead{({\AA})} & \colhead{(m{\AA})} &\colhead{} & \colhead{({\AA})} & \colhead{(m{\AA})}
}
\startdata
Be {\small II}	&	3130.42	&	85	$\pm $	12	&	Ti {\small II}	&	3341.88	&	22	$\pm $	3	\\
Be {\small II}	&	3131.07	&	49	$\pm $	10	&	Ti {\small II}	&	3349.04	&	31	$\pm $	3	\\
O {\small I} 	&	7771.94	&	283	$\pm $	8	&	Ti {\small II}	&	3349.41	&	62	$\pm $	3	\\
O {\small I} 	&	7774.16	&	194	$\pm $	14	&	Ti {\small II}	&	3361.22	&	53	$\pm $	3	\\
O {\small I} 	&	7775.39	&	172	$\pm $	21	&	Ti {\small II}	&	3372.80	&	45	$\pm $	3	\\
O {\small I} 	&	8447	&	348	$\pm $	24	&	Ti {\small II}	&	3383.77	&	29	$\pm $	4	\\
Mg {\small I}	&	3829.35	&	52	$\pm $	14	&	Ti {\small II}	&	3685.20	&	30	$\pm $	4	\\
Mg {\small I}	&	3832.30	&	156	$\pm $	6	&	Ti {\small II}	&	3759.29	&	24	$\pm $	4	\\
Mg {\small I}	&	3838.29	&	255	$\pm $	17	&	Ti {\small II}	&	3761.32	&	23	$\pm $	4	\\
Mg {\small I}	&	5167.32	&	25	$\pm $	7	&	Cr {\small II}	&	3120.36	&	50	$\pm $	4	\\
Mg {\small I}	&	5172.68	&	62	$\pm $	3	&	Cr {\small II}	&	3124.97	&	79	$\pm $	7	\\
Mg {\small I}	&	5183.60	&	85	$\pm $	9	&	Cr {\small II}	&	3132.05	&	89	$\pm $	16	\\
Mg {\small II}	&	4481	&	541	$\pm $	46	&	Cr {\small II}	&	3197.08	&	20	$\pm $	4	\\
Mg {\small II}	&	7877.05	&	137	$\pm $	60	&	Cr {\small II}	&	3358.49	&	22	$\pm $	3	\\
Mg {\small II}	&	7896.20	&	369	$\pm $	59	&	Cr {\small II}	&	3368.04	&	34	$\pm $	2	\\
Si {\small II}	&	3856.02	&	108	$\pm $	4	&	Cr {\small II}	&	3408.76	&	20	$\pm $	2	\\
Si {\small II}	&	3862.59	&	66	$\pm $	5	&	Cr {\small II}	&	3422.73	&	24	$\pm $	3	\\
Si {\small II}	&	4128.05	&	40	$\pm $	11	&	Mn {\small II}	&	3441.99	&	37	$\pm $	2	\\
Si {\small II}	&	4130.89	&	53	$\pm $	7	&	Mn {\small II}	&	3460.31	&	21	$\pm $	5	\\
Si {\small II}	&	5041.02	&	30	$\pm $	6	&	Fe {\small I}	&	3570.10	&	20	$\pm $	3	\\
Si {\small II}	&	5055.98	&	73	$\pm $	27	&	Fe {\small I}	&	3581.19	&	34	$\pm $	3	\\
Si {\small II}	&	6347.11	&	202	$\pm $	7	&	Fe {\small I}	&	3734.86	&	22	$\pm $	3	\\
Si {\small II}	&	6371.37	&	99	$\pm $	6	&	Fe {\small I}	&	3749.49	&	20	$\pm $	2	\\
Ca {\small II}	&	3158.87	&	306	$\pm $	14	&	Fe {\small II}	&	3154.20	&	54	$\pm $	7	\\
Ca {\small II}	&	3179.33	&	352	$\pm $	28	&	Fe {\small II}	&	3167.86	&	38	$\pm $	9	\\
Ca {\small II}	&	3181.28	&	45	$\pm $	4	&	Fe {\small II}	&	3177.53	&	25	$\pm $	5	\\
Ca {\small II}	&	3706.02	&	104	$\pm $	13	&	Fe {\small II}	&	3186.74	&	32	$\pm $	8	\\
Ca {\small II}	&	3736.90	&	178	$\pm $	7	&	Fe {\small II}	&	3192.91	&	24	$\pm $	9	\\
Ca {\small II}	&	3933.66	&	1409	$\pm $	94	&	Fe {\small II}	&	3193.80	&	41	$\pm $	10	\\
Ca {\small II}	&	3968.47	&	1053	$\pm $	156	&	Fe {\small II}	&	3196.07	&	30	$\pm $	6	\\
Ca {\small II}	&	8498.02	&	100	$\pm $	17	&	Fe {\small II}	&	3210.44	&	57	$\pm $	4	\\
Ca {\small II}	&	8542.09	&	587	$\pm $	34	&	Fe {\small II}	&	3213.31	&	84	$\pm $	5	\\
Ca {\small II}	&	8662.14	&	349	$\pm $	17	&	Fe {\small II}	&	3227.74	&	120	$\pm $	6	\\
Ti {\small II}	&	3190.87	&	22	$\pm $	6	&	Fe {\small II}	&	3247.18	&	20	$\pm $	7	\\
Ti {\small II}	&	3234.51	&	47	$\pm $	5	&	Fe {\small II}	&	3259.05	&	24	$\pm $	4	\\
Ti {\small II}	&	3236.57	&	28	$\pm $	3	&	Fe {\small II}	&	4923.92	&	27	$\pm $	9	\\
Ti {\small II}	&	3239.04	&	20	$\pm $	5	&	Fe {\small II}	&	5018.44	&	42	$\pm $	6	\\
Ti {\small II}	&	3241.98	&	26	$\pm $	6	&	Fe {\small II}	&	5169.03	&	66	$\pm $	4	\\
\enddata
\tablecomments{Observed lines with measured EW $>$ 20m{\AA}. Wavelengths are in air. }
\end{deluxetable}

\startlongtable
\begin{deluxetable}{lcc|lccl}
\tablecaption{Absorption Lines in GD 378 \label{tab:linelist_gd378}}
\tablehead{
\colhead{Ion} & \colhead{$\lambda$} & \colhead{EW} & \colhead{Ion} & \colhead{$\lambda$} & \colhead{EW} \\
\colhead{} & \colhead{({\AA})} & \colhead{(m{\AA})} &\colhead{} & \colhead{({\AA})} & \colhead{(m{\AA})}
}
\startdata
Be {\small II}	&	3130.42	&	8.6	$\pm $	1.8	&	Si {\small III}	&	1113.23	&	66	$\pm $	12	\\
Be {\small II}	&	3131.07	&	3.8	$\pm $	1.6	&	P {\small II}	&	1015.46	&	49	$\pm $	10	\\
C {\small II}	&	1009.86	&	22	$\pm $	7	&	P {\small II}	&	1154.00	&	38	$\pm $	13	\\
C {\small II}	&	1010.08	&	31	$\pm $	7	&	S {\small II}	&	1014.11	&	34	$\pm $	9	\\
C {\small II}	&	1010.37	&	63	$\pm $	12	&	S {\small II}	&	1014.44	&	63	$\pm $	9	\\
C {\small II}	&	1036.34	&	146	$\pm $	24	&	S {\small II}	&	1019.53	&	48	$\pm $	12	\\
C {\small II}	&	1037.02	&	236	$\pm $	57	&	S {\small II}	&	1124.99	&	45	$\pm $	15	\\
O {\small I} 	&	988.66	&	104	$\pm $	37	&	Ca {\small II}	&	3158.87	&	16	$\pm $	3	\\
O {\small I} 	&	988.77	&			{\it blended}	&	Ca {\small II}	&	3179.33	&	16	$\pm $	2	\\
O {\small I} 	&	990.13	&	124	$\pm $	47	&	Ca {\small II}	&	3736.90	&	5.6	$\pm $	0.6	\\
O {\small I} 	&	990.20	&			{\it blended}	&	Ca {\small II}	&	3933.66	&	165	$\pm $	2	\\
O {\small I} 	&	999.50	&	110	$\pm $	21	&	Ca {\small II}	&	3968.47	&	96	$\pm $	2	\\
O {\small I} 	&	1039.23	&	207	$\pm $	57	&	Ti {\small II}	&	3349.03	&	1.9	$\pm $	0.7	\\
O {\small I} 	&	1040.94	&	130	$\pm $	19	&	Ti {\small II}	&	3349.40	&	2.3	$\pm $	0.5	\\
O {\small I} 	&	1041.69	&	80	$\pm $	16	&	Cr {\small II}	&	3120.40	&	3.6	$\pm $	1.1	\\
O {\small I} 	&	1152.15	&	171	$\pm $	17	&	Cr {\small II}	&	3124.97	&	4.7	$\pm $	1.2	\\
O {\small I} 	&	7771.94	&	46	$\pm $	8	&	Cr {\small II}	&	3132.05	&	3.6	$\pm $	0.8	\\
O {\small I} 	&	7774.16	&	35	$\pm $	14	&	Mn {\small II}	&	3441.99	&	1.3	$\pm $	0.4	\\
O {\small I} 	&	7775.39	&	19	$\pm $	5	&	Fe {\small II}	&	1063.18	&	59	$\pm $	16	\\
O {\small I} 	&	8447	&	142	$\pm $	24	&	Fe {\small II}	&	1068.35	&	31	$\pm $	10	\\
Mg {\small I}	&	3838.29	&	8.6	$\pm $	2.0	&	Fe {\small II}	&	1144.94	&	63	$\pm $	12	\\
Mg {\small II}	&	4481	&	62	$\pm $	14	&	Fe {\small II}	&	1148.28	&	60	$\pm $	20	\\
Si {\small II}	&	992.68	&	355	$\pm $	39	&	Fe {\small II}	&	3154.20	&	8.3	$\pm $	1.6	\\
Si {\small II}	&	992.70	&			{\it blended}	&	Fe {\small II}	&	3167.86	&	3.2	$\pm $	1.0	\\
Si {\small II}	&	1020.70	&	98	$\pm $	21	&	Fe {\small II}	&	3210.44	&	4.2	$\pm $	0.7	\\
Si {\small II}	&	3856.02	&	10.7	$\pm $	1.8	&	Fe {\small II}	&	3213.31	&	7.5	$\pm $	1.6	\\
Si {\small II}	&	3862.59	&	5.8	$\pm $	0.9	&	Fe {\small II}	&	3227.74	&	12	$\pm $	2.0	\\
Si {\small II}	&	4130.89	&	8.4	$\pm $	2.7	&	Fe {\small II}	&	5169.03	&	7.7	$\pm $	1.9	\\
Si {\small II}	&	6347.11	&	37	$\pm $	7	&	Fe {\small III}	&	1122.52	&	42	$\pm $	8	\\
Si {\small II}	&	6371.37	&	12	$\pm $	2	&	Fe {\small III}	&	1124.88	&	52	$\pm $	13	\\
Si {\small III}	&	1109.97	&	31	$\pm $	12	&	Fe {\small III}	&	1126.729	&	50	$\pm $	24	\\
\enddata
\tablecomments{Si {\small II} $\lambda$989.87 and Fe {\small II} $\lambda$989.90 are blended with a combined EW of $\simeq$200 m{\AA}; similarly O {\small I} $\lambda$990.80 and Fe {\small II} $\lambda$990.86 have a blended EW of $\simeq$180 m{\AA}. These transitions are not used in the abundance analysis. Wavelengths are in vacuum below 3000\,{\AA}, and air above 3000\,{\AA}. }
\end{deluxetable}

\clearpage

\begin{figure}
\includegraphics[width=85mm]{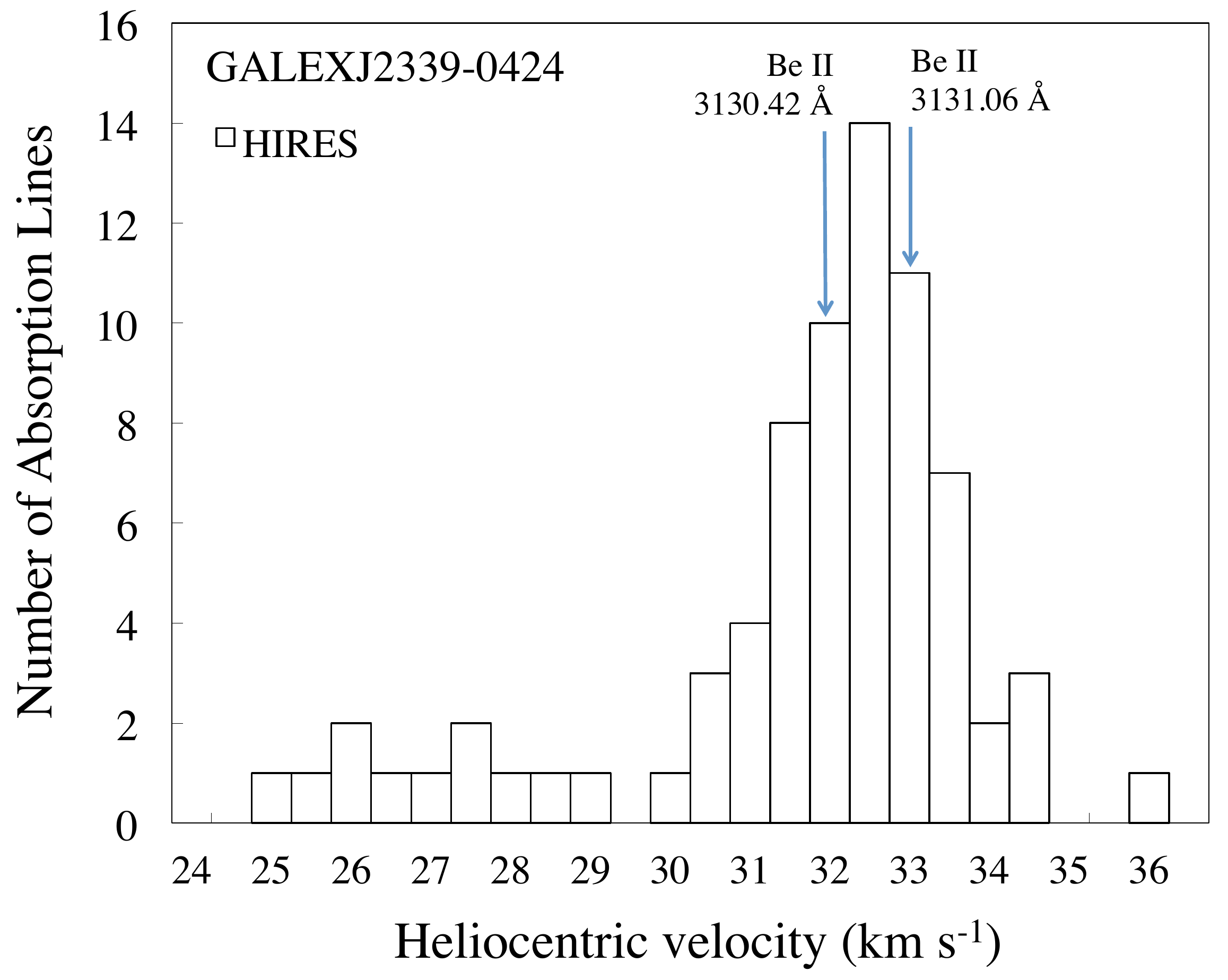}
\caption{Heliocentric radial velocities of absorption lines from Table \ref{tab:linelist_j2339}. The Be lines have velocities consistent with the other heavy element lines observed from the WD photosphere and thus are photospheric.  The low velocity tail between 25 to 28 km s$^{-1}$ comes from the O {\small I} lines 7772/7774/7775/8447\,{\AA}, the Mg {\small I} triplet 5167/5173/5184\,{\AA}, and the doublet of Si {\small II} at 6347/6371\,{\AA}.  See text for additional comments on these somewhat shifted RV lines. 
}
\label{fig:RV_hist_2339}
\end{figure}

\begin{figure}
\includegraphics[width=85mm]{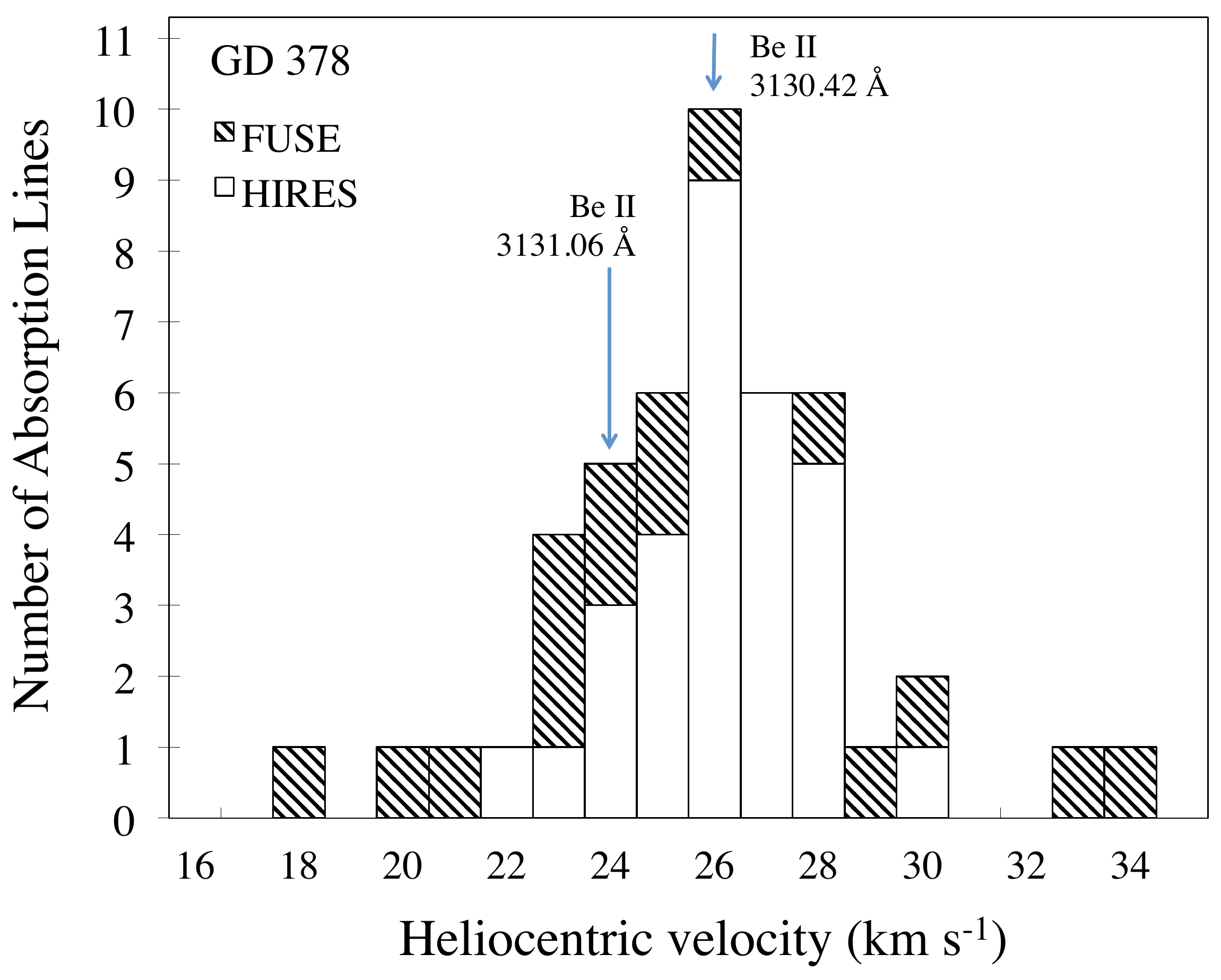}
\caption{Heliocentric radial velocities of absorption lines from Table \ref{tab:linelist_gd378}. Similar to GALEXJ2339, the Be lines in GD 378 are clearly photospheric.  FUSE RVs are only from the spectral range $\lambda$ $<$ 1100\,{\AA}, as described in the text.        
}
\label{fig:RV_hist_gd378}
\end{figure}

\section{Spectral Measurements}
\subsection{Absorption Lines}
Through all these observations, spectra rich with absorption lines appear in both stars. Line lists are given in Tables \ref{tab:linelist_j2339} and \ref{tab:linelist_gd378}, including the detections of the Be {\small II} resonance doublet at 3130.42 and 3131.06\,{\AA}, shown in Figures \ref{fig:Be_2339} and \ref{fig:Be_gd378}. 
These are the only observable optical beryllium lines $-$ the same ones used to obtain Be abundances in main sequence stars \citep{boesgaard1976be}, giant stars \citep{boesgaard2020}, and the Sun \citep{chmielewski1975}. Unlike main sequence stars, the Be lines in these WD spectra are almost entirely free from blending with lines of other elements, so we can be confident that our measured line strengths and derived abundances are not confused. The Be doublet lines arise out of the ground state, but to date no detection from the interstellar medium (ISM) has been made with a strict upper limit of $(^9{\rm Be/H}) \leq 7$ x $10^{-13}$ \citep{hebrard1997}.  Given these prior ISM studies, and that our measured radial velocities (RVs) are in excellent agreement with absorption lines of other elements (Figures \ref{fig:RV_hist_2339} and \ref{fig:RV_hist_gd378}),  we conclude that the origin of the Be lines is photospheric in the two WDs. 

EWs are measured with the IRAF task, {\it splot}, by fitting a Voigt function to the line profiles, and also by direct flux summation for unblended lines. The quoted values in Tables \ref{tab:linelist_j2339} and \ref{tab:linelist_gd378} are calculated from an average of three separate measurements with different continuum ranges.  For the uncertainties we combine the standard deviation of the three EW measurements with the average {\it splot} profile-fitting uncertainty in quadrature.

\subsection{Radial Velocities}
Figures \ref{fig:RV_hist_2339} and \ref{fig:RV_hist_gd378} show histograms of measured RVs for the sets of absorption lines for each star.  The plotted velocities, relative to the Sun, are not adjusted for the gravitational redshifts of the WDs, so the RVs of the photospheric absorption lines represent the WDs’ space motion (kinematic velocity) plus the gravitational redshift.  The primary take-away point from these plots is that, in both stars, the Be lines are consistent with coming from the WD photospheres, but some additional observations are noted as follows.

GALEXJ2339's RV distribution has a modestly blue-shifted tail, composed of lines of O {\small I}, Mg {\small I} and Si {\small II}, that must be photospheric due to the high values of the lower energy levels of the transitions. Such particular behavior of specific elements/species has been noted previously from HIRES spectra of the polluted WD PG 1225$-$079 \citep[in that case from Fe {\small I};][Figure 14]{klein2011}. Similar to the RV differential in GALEXJ2339, the blue-shifted lines in PG 1225$-$079 are also offset by about 6 km s$^{-1}$ from the peak of the RV distribution of the other WD absorption lines. \citet{klein2011} referenced the possible effects of Stark shifts by \citet{vennes2011}, but further investigation into the nature of these observations is beyond the scope of this work.

From GALEXJ2339's main peak distribution (RVs $>$ 28 km s$^{-1}$), the average is 32.3 km s$^{-1}$ with a standard deviation of 1.5 km s$^{-1}$.  Since the measured lines for this star come exclusively from HIRES, this standard deviation represents a reasonable uncertainty in measured RVs from HIRES data with setup as described in Section \ref{sec:hires_j2339}.  Based on measurements of RV calibration stars, we add an additional 1 km s$^{-1}$ uncertainty in the absolute scale. Using the above average RV and uncertainties, along with the gravitational redshift given in Table \ref{tab:params}, we find a heliocentric kinematic velocity for GALEXJ2339 of +6 $\pm$ 5 km s$^{-1}$.

For GD 378, RVs from HIRES and FUSE data are compared. However, we do not use the RVs of the eight photospheric lines from the region $\lambda$ $>$ 1100\,{\AA} of the FUSE spectrum, as those are all systematically offset from the main distribution, most likely due to differing wavelength calibrations of different spectroscopic segments of FUSE.  Otherwise, the average photospheric RVs from FUSE and HIRES separately are in very good agreement at 25.4 and 26.1 km s$^{-1}$, respectively.  The standard deviation from HIRES-only data is 1.7 km s$^{-1}$ (similar to GALEXJ2339 and representative of the typical relative measurement uncertainty from such HIRES data), while relative RV uncertainties from FUSE spectra are estimated to be $\sim$ 5-6 km s$^{-1}$ \citep[e.g.][]{moos2002, barstow2010}. Using the more precise and accurate HIRES-only measurements for the WD total heliocentric velocity, and the gravitational redshift from Table \ref{tab:params}, the kinematic velocity of GD 378 is -0.3 $\pm$ 4 km s$^{-1}$.

\subsection{Non-photospheric Lines \label{sec:non-phot}}
In both stars there are observed lines which arise from the ground state and have measured RVs that significantly disagree with the RVs of the photospheric lines.  We consider if their origins may be interstellar or circumstellar.

The Na {\small I} D resonance doublet ($\lambda$5889.95/5895.92\,{\AA}) is observed in GALEXJ2339 at an RV of $-$7 $\pm$ 1.5 km s$^{-1}$, i.e.~blue-shifted by 39 km s$^{-1}$ from the photospheric average of 32.3 $\pm$ 1.5 km s$^{-1}$, and thus clearly incompatible with the WD photosphere. These lines also do not agree very well with the kinematic (circumstellar) velocity of +6 $\pm$ 5 km s$^{-1}$ within 1-2$\sigma$ uncertainties, but could be considered in agreement at the 3$\sigma$ level.   Absorption from the intervening ISM may be a more likely source in this case, as the possible similarity with a line-of-sight ISM cloud velocity\footnote{\url http://lism.wesleyan.edu/LISMdynamics.html} \citep{redfieldlinsky2008} would support an ISM origin. Together these observations suggest that the Na {\small I} D lines in GALEXJ2339 are probably formed in the ISM, but we do not rule out the possibility of a circumstellar origin.

In GD 378 non-photospheric components of C {\small II} 1036.34, Ca {\small II} K ($\lambda$3933.66\,{\AA}), and O {\small I} 1039.23\,{\AA} are well separated in RV from the photospheric components of these lines and can even be seen resolved in the spectra presented in Section \ref{sec:abund}.  Along with Ca {\small II} H ($\lambda$3968.47\,{\AA}) and the N {\small I} triplet 1134.17/1134.42/1134.98\,{\AA}, this set of lines are all observed at RVs around $\sim$$-$20 km s$^{-1}$, which is neither consistent with the WD photosphere (26 $\pm$ 3 km s$^{-1}$) nor its kinematic (circumstellar) velocity (-0.3 $\pm$ 4 km s$^{-1}$), but however does agree well with at least one known line-of-sight cloud velocity \citep{redfieldlinsky2008}.  Thus the features are almost certainly interstellar.

 \begin{table}
\caption{WD Parameters \label{tab:params}}
\begin{center}
\begin{tabular}{lrrl}
\hline 
\hline
Parameter 	&	J2339$-$0424	& GD 378 	\\
\hline													
G (mag)  &  16.2  & 14.3 \\
Distance (pc)  & 90  & 44 \\
$T_{\rm eff}$ (K)  & 13735 (500) & 15620 (500)\\
log $g$  & 7.93 (0.09) & 7.93 (0.06)  \\
$M_{\rm WD} (M_\sun$)  &   0.548 (0.051) & 0.551 (0.031) \\   
$R_{\rm WD} (R_\sun$)  &   0.0133 (0.0008) & 0.0133 (0.0005) \\    
Grav.~redshift (km s$^{-1}$)  &  26.2 (4.0) & 26.4 (2.5) \\  
Cooling age (Myr) & 241 (6) & 157 (3) \\    
log ($M_{\rm CVZ}/M_{\rm WD}$) & $-$5.29 (0.30) & $-$5.77 (0.25) \\   
$\dot{M}$ (g s$^{-1}$)   &  1.7 x 10$^9$  & 1.8 x 10$^8$  \\
\hline
\end{tabular}
\end{center}
\tablecomments{Gmag and distance (inverse parallax) are from Gaia DR2.  $M_{\rm WD}$, $R_{\rm WD}$, gravitational redshift, cooling age, and $M_{\rm CVZ}$ (CVZ = convection zone) are from the Montreal White Dwarf Database \citep[MWDD;][]{dufourMWDD}\footnote{\url{http://dev.montrealwhitedwarfdatabase.org/evolution.html}}. Uncertainties given in parentheses represent the range in values for each parameter considered at the upper and lower limits of the $T_{\rm eff}$/\,log $g$ models (as described in Section \ref{sec:model}). $\dot{M}$ is the mass flow rate (see Section \ref{sec:abund}). }
\end{table}

\section{Model Atmospheres \label{sec:model}}
In He-atmosphere WDs with temperatures $\lesssim$20,000\,K, the presence of hydrogen and other elements can have a non-negligible effect on the atmosphere structure due to the additional opacity from these pollutants \citep{dufour2007, dufour2010, coutu2019}.   Taking this into account, here we use the same
methods and codes as described in \citet{dufour2007} and \citet{blouin2018a}, briefly summarized as follows. We started by fitting Sloan Digital Sky Survey \citep[SDSS;][]{alam2015} {\it ugriz} photometry\footnote{We apply the SDSS-to-AB magnitude corrections given in \citet{eisenstein2006}.} and Gaia parallax simultaneously with the Ca {\small II} H\&K region and H$\alpha$ from low resolution spectra.  For GALEXJ2339 we use the Kast spectrum described in Section \ref{sec:kast_j2339}, and for GD 378 we use the spectrum obtained by \citet{bergeron2011}, described therein.  This provides a first estimate of the effective temperature ($T_{\rm eff}$), gravity (\,log $g$), hydrogen abundance [H/He] ($\equiv$ log $n$(H)/$n$(He) ), and overall heavy element presence through [Ca/He] ($\equiv$ log $n$(Ca)/$n$(He) ), where all other elements up to Sr are included scaled to Ca in CI chondrite proportions \citep{lodders2003}. No de-reddening corrections were applied since it is not expected to be significant for stars within 100 pc.  

Next we compute an atmospheric structure using the above parameters and
from it, grids of synthetic spectra for each element, which we interpolate to fit the abundances.  With those grids, we run a fit of the HIRES data to obtain a first
estimation of the abundances of all detected elements. We then recalculate the structure using these estimated
abundances and repeat the fitting as many times as necessary until a stable solution is found. 
From this procedure we set our best-fit estimation of the nominal parameters for $T_{\rm eff}$ and log $g$ (given in Table \ref{tab:params}).  For both stars these atmospheric parameters correspond to WD masses of 0.55\,$M_\sun$, which according to an initial-final mass relationship for WDs \citep[Figure 5,][]{cummings2018}, indicate their progenitor star masses were $\simeq$1\,$M_\sun$, probably G-type stars.

 We explored uncertainties in $T_{\rm eff}$ and log $g$ from the range of models that can fit within the error bars of the photometry fitting described above.  However, the SDSS error bars are so small in our stars ($<$ 1\%), that we found this method resulted in unreasonably small errors ($\simeq$100\,K) for the effective temperatures, which we know $-$ from different modeling methods in the literature, and the changes seen over time from new developments in model structures, etc. $-$ can have much larger uncertainties.  Thus, we do not attempt to assign fitting errors to the atmospheric parameters.  Instead we assume more typical uncertainties ($\sim$3\% of the $T_{\rm eff}$ value) for helium-dominated WDs in this $T_{\rm eff}$ range \citep{bergeron2011, koesterkepler2015}, choosing $\pm$ 500 K (to be conservative), with corresponding log $g$ values consistent with fitting the photometry.  Note that due to Gaia parallax constraints, uncertainties in log $g$ are negligible.  This results in lower and upper  $T_{\rm eff}$/\,log $g$ limits for GALEXJ2339 of 13235/7.84 and 14235/8.02, and for GD 378 of 15120/7.87 and 16120/7.98.  See Appendix A for details on how we apply these limits.

\begin{figure}
\includegraphics[width=85mm]{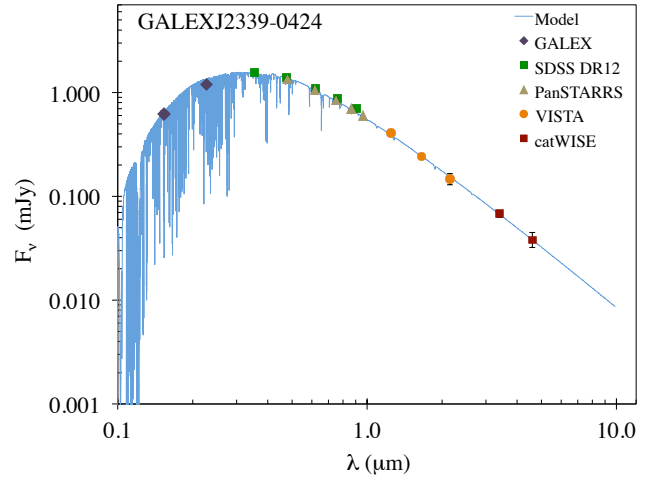}
\caption{Spectral energy distribution for GALEXJ2339.}
\label{fig:j2339_sed}
\end{figure}

\begin{figure}
\includegraphics[width=85mm]{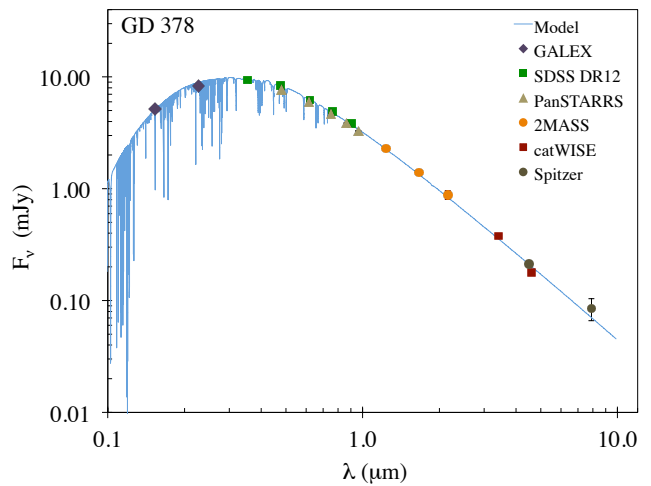}
\caption{Spectral energy distribution for GD 378.}
\label{fig:gd378_sed}
\end{figure}

\begin{table*}
\caption{Steady State Beryllium ratios relative to CI Chondrites \label{tab:ratios_per_ci}}
\begin{center}
\begin{tabular}{lrrrrrlrr}
\hline 
\hline
WD name	&	Mg/Fe	&		Be/O	&		Be/Mg	&		Be/Si	&		Be/Fe	\\
\hline															
GALEXJ2339	&	1.4	&		121	&		190	&		191	&		267	\\
GD 378	&	0.7	&		36	&		123	&		128	&		80	\\
\hline															
PG 1225-079	&	0.6	&		\ldots	&	$<$	33	&	$<$	38	&	$<$	20	\\
GD 362	&	0.2	&		\ldots	&	$<$	23	&	$<$	16	&	$<$	5	\\
SDSSJ1242+5226	&	1.4	&	$<$	0.9	&	$<$	6	&	$<$	4	&	$<$	8	\\
SDSSJ0738+1835	&	1.0	&	$<$	6	&	$<$	6	&	$<$	9	&	$<$	6	\\
G200-39	&	0.5	&	$<$	187	&	$<$	898	&	$<$	639	&	$<$	481	\\
Ton 345	&	0.6	&	$<$	122	&	$<$	41	&	$<$	27	&	$<$	26	\\
\hline															
Sun	&	1.02	&		0.54	&		1.06	&		1.16	&		1.08	\\
F \& G stars (avg)	&	1.3	&		0.4	&		0.9	&		1.1	&		1.2	\\
Bulk Earth	&	1.07	&		2.9	&		1.26	&		1.32	&		1.35	\\
\hline
\end{tabular}
\end{center}
\tablecomments{Ratios are by number and relative to CI chondrite ratios from \citet{lodders2020}. Abundances for GALEXJ2339 and GD 378 are from Tables \ref{tab:2339_abund} and \ref{tab:gd378_abund}, respectively, and comparison WD abundances are from Table \ref{tab:compare_abund}. Values for the Sun are from \citet{lodders2020}; F \& G stars from \citet{boesgaard1976be} and \citet{reddy2003}; bulk Earth from \citet{allegre2001}. The derived limits for G200-39 are not very restrictive, but they are included here as this WD accreted a Kuiper Belt analog, and we are showing that the upper limits do not preclude a beryllium over-abundance in such an object. For the WDs we adopt the steady state values using diffusion timescales from the Montreal White Dwarf Database \citep[MWDD;][]{dufourMWDD}.  Typical uncertainties on the measured Be ratios are about 50\%.   Note that both GALEXJ2339 and GD 378 have oxygen excesses as described in Section \ref{sec:abund}, thus the Be/O ratios are somewhat less dramatic than those relative to Mg, Si and Fe. If any WD system happens to be in an increasing phase of accretion, then all its tabulated WD abundance ratios or upper limits would be larger by up to factors of two (see text Section \ref{sec:accdiff}).}
\end{table*}

Spectral energy distributions are given in Figures \ref{fig:j2339_sed} and \ref{fig:gd378_sed} with available photometry and our best-fit models for each star plotted in blue.  The photometric data come from GALEX \citep{bianchi2017}, SDSS \citep{alam2015}, Pan-STARRS \citep{flewelling2020}, 2MASS \citep{cutri2003}, VISTA \citep{mcmahon2013}, and catWISE \citep{eisenhardt2020} surveys.  Spitzer fluxes for GD 378 are from \citep{mullally2007}. We checked the Spitzer archive at the position of GALEXJ2339, which at first returned a positive result suggesting that the WD may have been (unintentionally) observed in a prior field observation. Unfortunately, it turned out that GALEXJ2339 was just outside the imaged field of view.  From Figs  \ref{fig:j2339_sed} and \ref{fig:gd378_sed} we see that with the available data, neither WD displays evidence for an infrared excess due to circumstellar dust.  Although early on it was recognized that heavy pollution and infrared excess are correlated \citep{jura2008, farihi2009}, more recently it has been shown that only one in 30 polluted WDs exhibit an infrared excess when examined with Spitzer's 3$-$4$\micron$ IRAC photometry \citep{wilson2019irxs}.  Even heavily polluted WDs do not always have detected infrared excesses \citep[e.g.][]{klein2011, raddi2015, xu2017, hoskin2020}.

UV wavelengths are the most sensitive to the effects of interstellar reddening and line blanketing from heavy element pollution, both of which could cause the measured UV flux to be lower than predicted by a model without accounting for these factors.  On the other hand if measured UV flux is much higher than the model, that would suggest a higher temperature model is needed to match the full SED. Both WDs in this study have GALEX photometry. As shown in Figures \ref{fig:j2339_sed} and \ref{fig:gd378_sed}, our model spectra are well-matched with the GALEX points of both stars, which lends support to the values derived from our model $T_{\rm eff}$/\,log $g$ fits with the assumption of negligible reddening.

\section{abundances \label{sec:abund}}

\begin{figure*}
\begin{center}
\includegraphics[width=140mm]{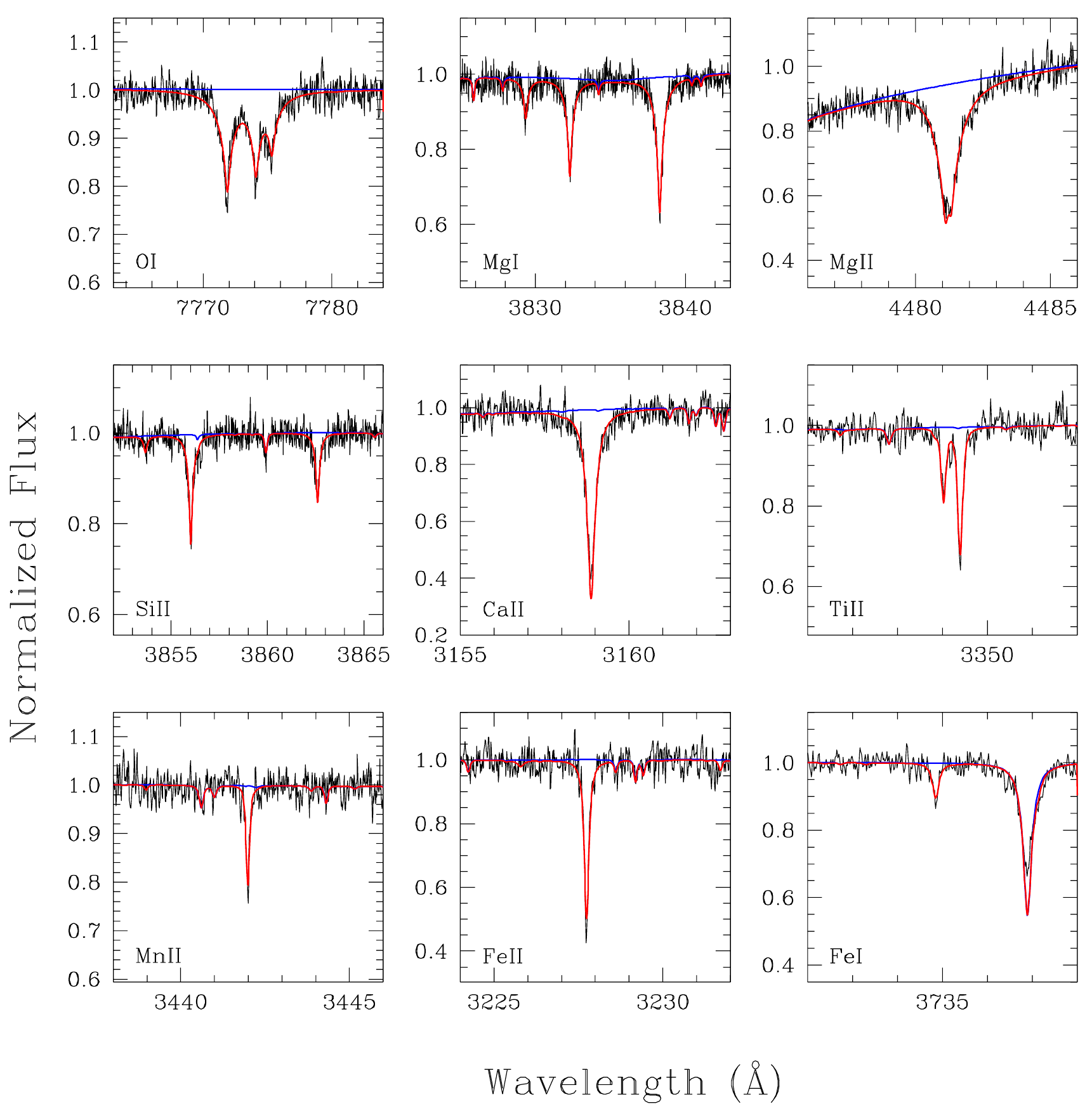}
\caption{Portions of the Keck/HIRES spectrum of GALEXJ2339, displaying examples of each of the detected elements (along with Be from Figure \ref{fig:Be_2339}).  Wavelengths are in air and shifted to the laboratory frame of rest.  The red line is our best-fit model, and the blue line is the same model with the abundance of the indicated element set to zero.  In the lower right panel of Fe {\small I}, the stronger absorption line at 3736.9\,{\AA} is from Ca {\small II}.  }
\label{fig:j2339_spectra}
\end{center}
\end{figure*}

Our most noteworthy finding is that GALEXJ2339 and GD 378 have extraordinarily high Be abundances (relative to other rock-forming elements) compared with cosmic abundances and other heavily polluted WDs, as shown in Table \ref{tab:ratios_per_ci}.  The full set of measured averaged abundances for all detected elements are given in Tables \ref{tab:2339_abund} and \ref{tab:gd378_abund}, with example model fits to portions of the spectra shown in Figure \ref{fig:j2339_spectra} for GALEXJ2339, and Figures \ref{fig:gd378_spectra1} and \ref{fig:gd378_spectra2} for GD 378. Abundances for major elements and upper limits for Be in the comparison WDs are given in Table \ref{tab:compare_abund}.  Details of our abundance fitting procedures are described in Appendix A.

\begin{figure*}
\begin{center}
\includegraphics[width=140mm]{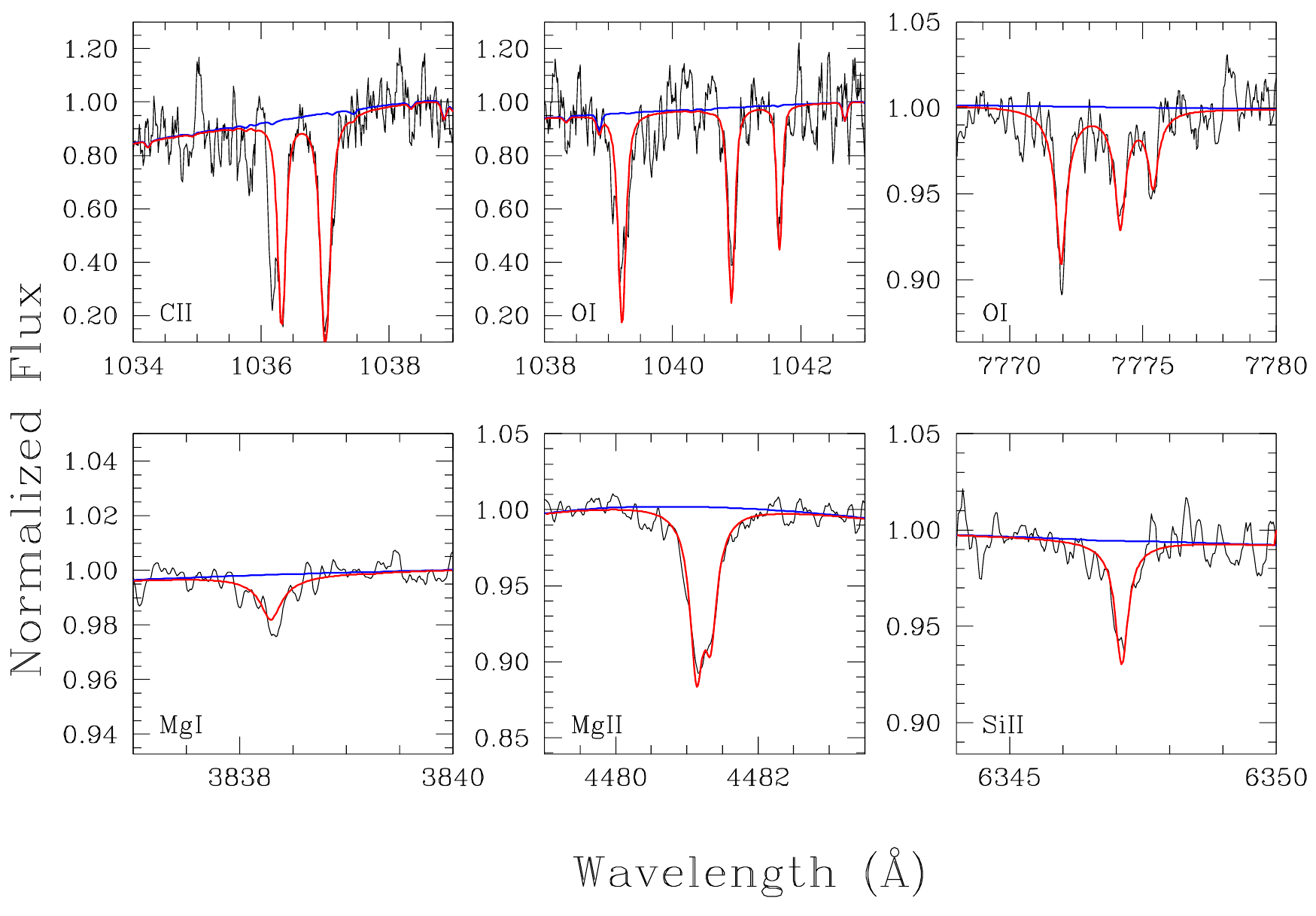}
\caption{Portions of the FUSE (wavelengths less than 3000\,{\AA} in vacuum) and Keck/HIRES (wavelengths greater than 3000\,{\AA} in air) spectra of GD 378, displaying examples of each of the lighter detected elements up through Si (along with Be from Figure \ref{fig:Be_gd378}). The data are smoothed by a 5-point average. The red and blue model lines have the same meaning as in Figure \ref{fig:j2339_spectra}. Non-photospheric components of C {\small II} 1036.34\,{\AA} and O {\small I} 1039.23\,{\AA} are present, blue-shifted from the photospheric lines. These features are almost certainly interstellar (see Section \ref{sec:non-phot}). }
\label{fig:gd378_spectra1}
\end{center}
\end{figure*}

\begin{figure*}
\begin{center}
\includegraphics[width=140mm]{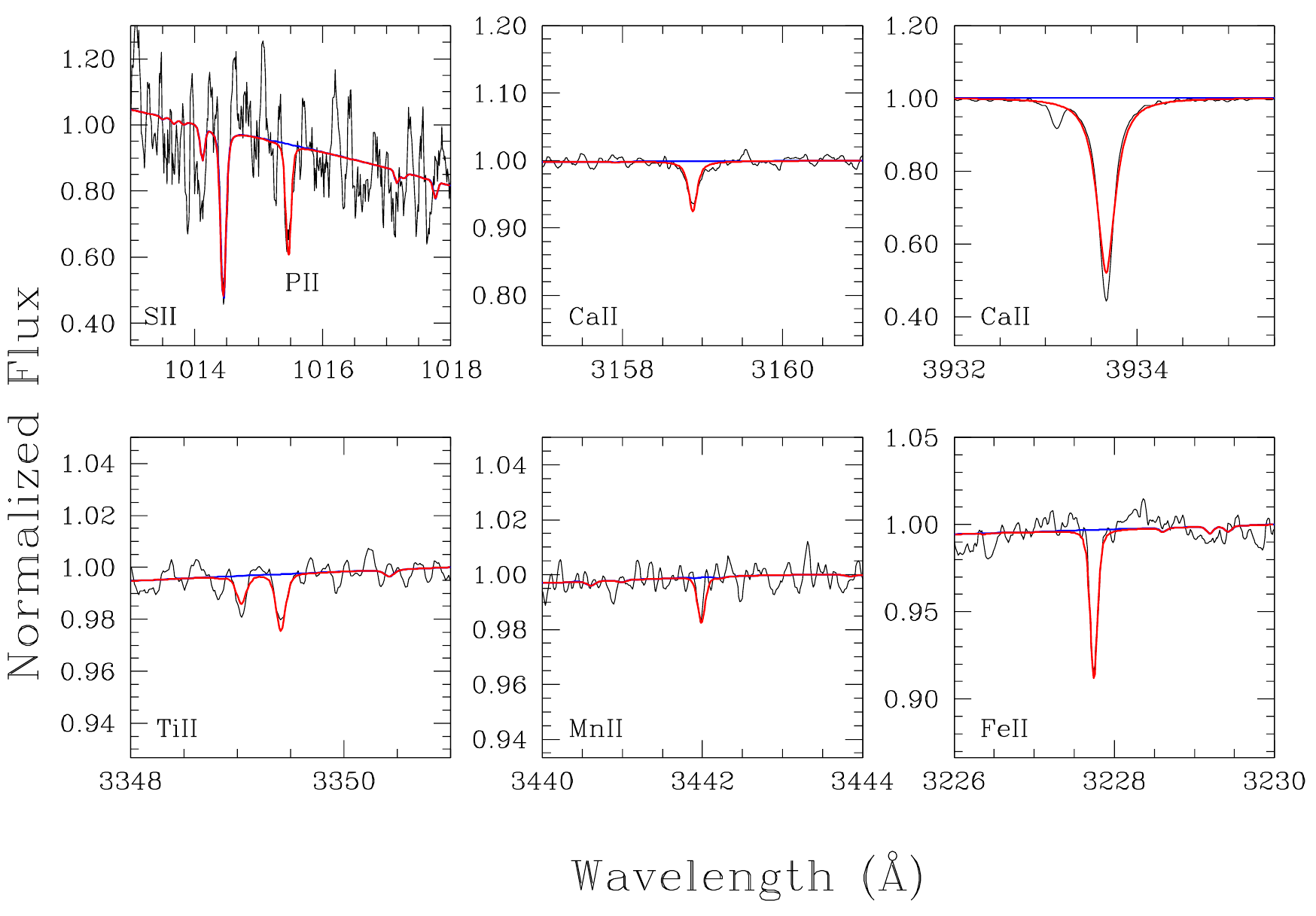}
\caption{Continuation of Figure \ref{fig:gd378_spectra1} for GD 378 elements P and heavier.  Phosphorous and sulfur appear together in the upper left panel, where the stronger line at 1014.4\,{\AA} is from S {\small II} and the line at 1015.5\,{\AA} is from P {\small II}. The absorption feature at 3933.1\,{\AA} is a blue-shifted (most likely interstellar) component of the Ca {\small II} K-line (see Section \ref{sec:non-phot}).  }
\label{fig:gd378_spectra2}
\end{center}
\end{figure*}

\begin{figure}
\begin{center}
\includegraphics[width=85mm]{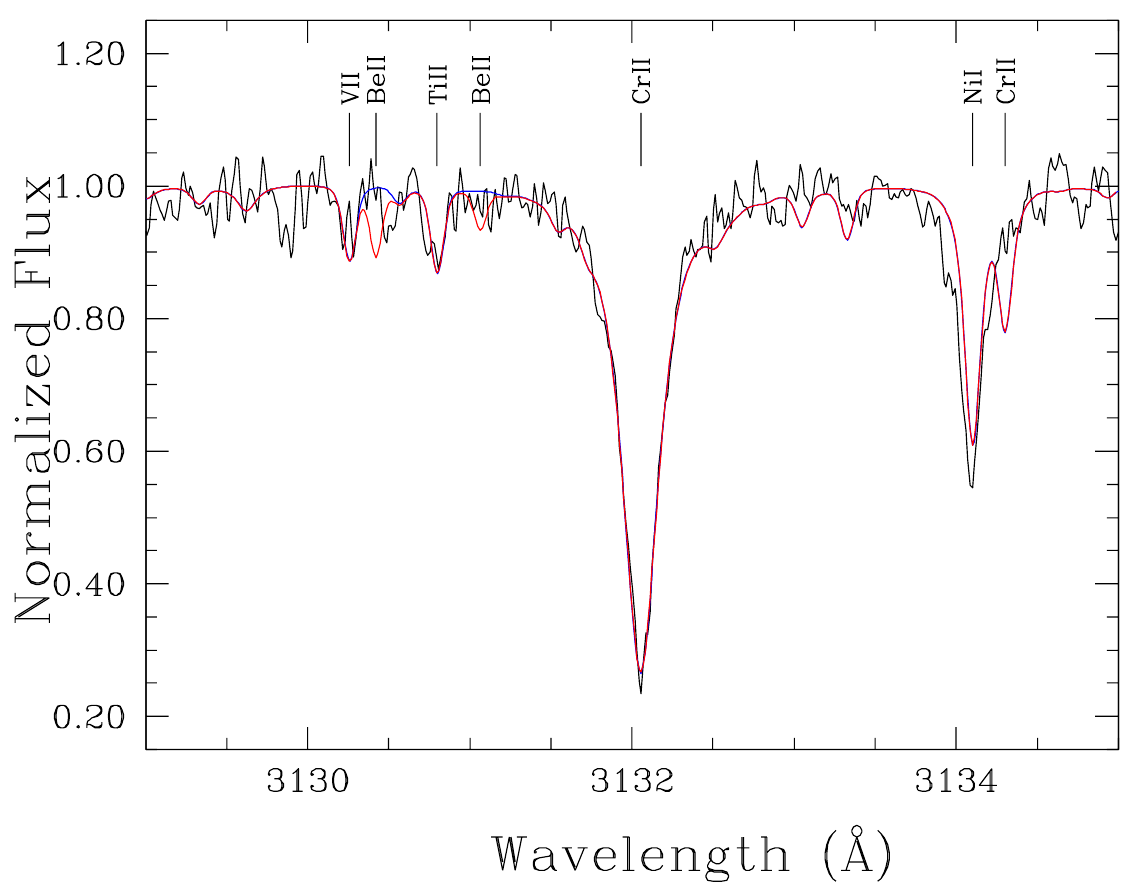}
\caption{Be {\small II} region in GD 362, similar to figure \ref{fig:Be_2339}, but here no features coincident with the Be lines are detected.  The red model line shows the Be upper abundance limit from Table \ref{tab:compare_abund}.}
\label{fig:Be_gd362}
\end{center}
\end{figure}

For the comparison WDs, we re-examined a sample of heavily polluted WDs for which the authors have obtained high quality HIRES spectra covering the 3130 {\AA} region of the Be doublet.  For most of these we used previously published major element abundances and atmospheric parameters to derive Be upper limits from the spectra (see references in Table \ref{tab:compare_abund}).  However for two stars, PG1225$-$079 and SDSSJ1242+5226, we refit the atmospheric parameters with the inclusion of all elements in the atmospheric structure calculations.  This resulted in lower $T_{\rm eff}$ by 1000\,K for PG1225 and 2300\,K for SDSSJ1242+5226 compared to previous studies \citep{klein2011, raddi2015}, and correspondingly altered abundance ratios. Since we only have a blue HIRES spectrum available for SDSSJ1242+5226, which doesn't cover the OI 7772 lines modeled by \citet{raddi2015}, we roughly estimated oxygen to have the same reduced abundance as other major elements compared to the analysis by \citep{raddi2015}. These changes in absolute abundances of a factor of two or three have very little effect on the relative abundances between elements \citep{klein2010}, and are not important in our overall comparisons regarding the abundance of Be.

Figure \ref{fig:Be_gd362} shows our model upper limit on the detection of the Be lines in the extremely heavily polluted WD, GD 362, as an example of how we derive the Be abundance upper limits for WDs in Table \ref{tab:compare_abund}.  Figure \ref{fig:Be_gd362}, Table \ref{tab:ratios_per_ci}, and panel (A) of Figure \ref{fig:ratios} demonstrate that the detections of Be in GALEXJ2339 and GD 378 are possible only because of the dramatic over-abundance of Be in these stars, while a ``normal'' (chondritic) Be abundance is below our detection limit in even the most heavily polluted WDs with good SNR.

The masses of accreted material in the convection zones (CVZs) are calculated using Table \ref{tab:params} parameters ($M_{\rm WD}$ and fractional mass of the CVZ) and the measured abundance ratios for each element. Mass flux (flow rates), $\dot{M}$ (g s$^{-1}$), can then be derived using the CVZ pollution masses divided by the settling times from Tables \ref{tab:2339_abund} and \ref{tab:gd378_abund}.   If accretion is ongoing the calculated mass flux represents the accretion rate of material into the WD atmosphere. After accretion ends in WDs with relatively long settling times, heavy elements can continue to be observable in the WD photosphere for some time before they diffuse down out of sight (more on this in Section \ref{sec:accdiff}).  In that situation, the mass flux can be more accurately thought of as the diffusion flux out of the base of the CVZ. Whatever the case, the total mass flow rates of  $\dot{M}$ = 1.7 x 10$^9$ g s$^{-1}$ for GALEXJ2339 and $\dot{M}$ = 1.8 x 10$^8$ g s$^{-1}$ for GD 378 are normal for polluted WDs \citep[compare Figure 3 of][]{xudusty2019}.

Summing up the observed mass from elements heavier than He, we find the minimum masses of the parent bodies to be 9.4 x 10$^{22}$ g for GALEXJ2339 and 4.5 x 10$^{21}$ g for GD 378, comparable to some of the most massive solar system asteroids: 10 Hygiea (fourth most massive) and  7 Iris (16th most massive), respectively.  The actual masses of the parent bodies could be much larger if accretion has been going on for more than a few settling times.  These stars also have a lot of hydrogen, 4.4 x 10$^{23}$ g  and 7.7 x 10$^{21}$ g for GALEXJ2339 and GD 378, respectively, but we  don't know for sure how much H is associated with the current pollution, since unlike the elements heavier than He, H tends to float to the top of the atmosphere, and accumulates from all accretion over time. We do know that the large amount of H in GALEXJ2339 is over the limit of the amount that could be primordial and preceding the DA-to-DB transition (otherwise the WD would never appear as spectral type DB dominated by optical He {\small I} lines) according to \citet[][Figure 4]{rolland2020}.  This implies that most of the H must have been accreted since that transition time ($\sim$180 Myr ago for GALEXJ2339).  The accretion of water-ice-rich bodies is the most likely explanation in such polluted WD systems with large amounts of H \citep{farihi2013, raddi2015, gentilefusillo2017, hoskin2020}, consistent with our composition analysis regarding oxygen excess in the two WDs studied here (Section \ref{sec:oexcess} below).

\begin{deluxetable*}{lrrrrcccccc}
\tablecaption{GALEXJ2339 Abundances and Parent Body Mass Compositions \label{tab:2339_abund}}
\tablewidth{0pt}
\tablehead{																								Z	&		log($n$(Z)/$n$(He))							&		$n$(Z)/$n$(He)	&	$\sigma_{\rm spread}$	&	$\sigma_{T_{\rm eff}}$	&		$n$(Z)/$n$(O)				&	$\tau$(Z)	&	\multicolumn{4}{c}{\% mass composition}							\\
\cline{8-11}																																	
	&									&		 ($10^{-8}$)	&	 ($10^{-8}$)	&	 ($10^{-8}$)	&		($10^{-3}$)				&	(Myr)	&	Steady	&	Rock only	&	Decreas-	&	CI	\\
	&									&			&		&		&						&		&	State	&	no H$_2$O	&	ing Phase	&	Chondrites	}
\startdata																																	
H	&	$	-3.51	^{+	0.18	}_{-	0.31	}	$	&		31260	&	14430	&	7057	&	$		\ldots		$	&	\ldots	&		&		&		&		\\
Be	&	$	-10.39	^{+	0.19	}_{-	0.34	}	$	&		0.0041	&	0.0010	&	0.0020	&	$	0.0136	\pm	0.0068	$	&	2.47	&	3.9E-04	&	7.0E-04	&	2.2E-06	&	2.2E-06	\\
O	&	$	-5.52	^{+	0.04	}_{-	0.05	}	$	&		298.9	&	25.7	&	18.6	&	$		\ldots		$	&	1.89	&	66.3	&	39.5	&	3.0	&	45.4	\\
Mg	&	$	-6.58	^{+	0.11	}_{-	0.14	}	$	&		26.4	&	4.0	&	6.0	&	$	88.4	\pm	21.3	$	&	1.90	&	8.8	&	15.9	&	0.37	&	9.5	\\
Si	&	$	-6.59	^{+	0.07	}_{-	0.08	}	$	&		25.6	&	4.3	&	1.6	&	$	85.8	\pm	16.0	$	&	1.90	&	9.9	&	17.9	&	0.43	&	10.8	\\
Ca	&	$	-8.03	^{+	0.26	}_{-	0.75	}	$	&		0.94	&	0.28	&	0.72	&	$	3.1	\pm	2.4	$	&	1.27	&	0.78	&	1.4	&	3.2	&	0.88	\\
Ti	&	$	-9.58	^{+	0.21	}_{-	0.40	}	$	&		0.027	&	0.007	&	0.015	&	$	0.089	\pm	0.049	$	&	1.15	&	0.029	&	0.052	&	0.52	&	0.045	\\
Cr	&	$	-8.73	^{+	0.16	}_{-	0.26	}	$	&		0.19	&	0.03	&	0.08	&	$	0.62	\pm	0.25	$	&	1.18	&	0.21	&	0.39	&	2.6	&	0.26	\\
Mn	&	$	-9.03	^{+	0.16	}_{-	0.25	}	$	&		0.094	&	0.006	&	0.041	&	$	0.32	\pm	0.12	$	&	1.17	&	0.12	&	0.21	&	1.4	&	0.19	\\
Fe	&	$	-6.99	^{+	0.18	}_{-	0.30	}	$	&		10.3	&	2.6	&	4.4	&	$	34.3	\pm	15.7	$	&	1.22	&	12.3	&	22.1	&	85.8	&	18.6	\\
\hline																																	
Li	&	$<$	-8.2							&	$<$	0.6	&		&		&	$<$	2.1				&	2.62	&	\ldots	&	\ldots	&	\ldots	&	1.5E-04	\\
Na	&	$<$	-8.0							&	$<$	1.0	&		&		&	$<$	3.3				&	1.82	&	\ldots	&	\ldots	&	\ldots	&	0.51	\\
Al	&	$<$	-7.7							&	$<$	2.2	&		&		&	$<$	7.5				&	1.85	&	0.74\tablenotemark{\dag}	&	1.32\tablenotemark{\dag}	&	$<$ 0.03	&	0.84	\\
V	&	$<$	-10.3							&	$<$	0.0046	&		&		&	$<$	0.015				&	1.14	&	\ldots	&	\ldots	&	\ldots	&	5.4E-03	\\
Ni	&	$<$	-8.0							&	$<$	1.0	&		&		&	$<$	3.3				&	1.28	&	0.72\tablenotemark{\dag}	&	1.30\tablenotemark{\dag}	&	$<$ 2.6	&	1.10	\\
\hline																									\enddata
\tablenotetext{\dag}{assumed contribution, see text Section \ref{sec:oexcess}}
\tablecomments{Abundances by number, n, and uncertainties, $\sigma$, as defined in Appendix A.  Upper limits are from non-detections of Al {\small I} 3962\,{\AA}, V {\small II} 3125\,{\AA}, and Ni {\small I} 3515\,{\AA}.   Uncertainties in Log abundances come from Equation \ref{eq:sigma_ZtoHe}, and for n(Z)/n(O) are calculated according to Equation \ref{eq:sigma_ZtoZ}.  $\tau$(Z) are the settling times in the WD atmosphere from the MWDD \citep{dufourMWDD}.  The compositions by mass in the four columns on the right are the following (see also Section \ref{sec:accdiff}): ``steady state'' = accretion-diffusion equilibrium ; ``Rock only'' = the steady state phase with `excess' O removed from the O abundance and attributed to water ice; ``Decreasing Phase'' = the decreasing phase after $\sim$ 7-8 Be settling times; ``CI chondrites'' = meteoritic abundances from \citet{lodders2020}.  
}																							
\end{deluxetable*}

\begin{deluxetable*}{lrrrrcccccc}
\tablecaption{GD 378 Abundances and Parent Body Mass Compositions \label{tab:gd378_abund}}
\tablewidth{0pt}
\tablehead{
Z	&		log(Z/He)							&		$n$(Z)/$n$(He)	&	$\sigma_{\rm spread}$	&	$\sigma_{T_{\rm eff}}$	&		$n$(Z)/$n$(O)				&	$\tau$(Z)	&	\multicolumn{4}{c}{\% mass composition}							\\
\cline{8-11}																																	
	&									&		 ($10^{-8}$)	&	 ($10^{-8}$)	&	 ($10^{-8}$)	&		($10^{-3}$)				&	(Myr)	&	Steady	&	Rock only	&	Decreas-	&	CI	\\
	&									&			&		&		&						&		&	State	&	no H$_2$O	&	ing Phase	&	Chondrites	}
\startdata																																	
H	&	$	-4.48	^{+	0.12	}_{-	0.17	}	$	&		3311	&		&	1091	&	$		\ldots		$	&	\ldots	&		&		&		&		\\
Be	&	$	-11.44	^{+	0.13	}_{-	0.19	}	$	&		0.00036	&	0.00007	&	0.00010	&	$	0.0040	\pm	0.0014	$	&	1.07	&	1.2E-04	&	2.5E-04	&	2.2E-06	&	2.2E-06	\\
C	&	$	-7.35	^{+	0.15	}_{-	0.24	}	$	&		4.48	&	1.58	&	1.04	&	$	49.2	\pm	23.3	$	&	0.91	&	2.4	&	4.9	&	0.12	&	4.1	\\
O	&	$	-6.04	^{+	0.18	}_{-	0.31	}	$	&		91.17	&	20.7	&	41.4	&			\ldots			&	0.84	&	70.3	&	39.5	&	6.3	&	45.4	\\
Mg	&	$	-7.44	^{+	0.14	}_{-	0.20	}	$	&		3.67	&	0.50	&	1.26	&	$	40.2	\pm	11.6	$	&	0.83	&	4.3	&	8.8	&	0.43	&	9.5	\\
Si	&	$	-7.49	^{+	0.09	}_{-	0.12	}	$	&		3.23	&	0.76	&	0.21	&	$	35.5	\pm	18.0	$	&	0.78	&	4.7	&	9.6	&	0.76	&	10.8	\\
P	&	$	-9.35	^{+	0.20	}_{-	0.39	}	$	&		0.044	&	0.016	&	0.021	&	$	0.49	\pm	0.2	$	&	0.68	&	0.081	&	0.17	&	0.044	&	0.10	\\
S	&	$	-7.81	^{+	0.19	}_{-	0.35	}	$	&		1.55	&	0.45	&	0.73	&	$	17.0	\pm	6.2	$	&	0.65	&	3.1	&	6.3	&	2.5	&	5.4	\\
Ca	&	$	-8.70	^{+	0.26	}_{-	0.76	}	$	&		0.20	&	0.04	&	0.16	&	$	2.2	\pm	1.0	$	&	0.55	&	0.59	&	1.2	&	3.0	&	0.88	\\
Ti	&	$	-10.13	^{+	0.22	}_{-	0.46	}	$	&		0.0073	&	0.0020	&	0.0044	&	$	0.081	\pm	0.03	$	&	0.52	&	0.028	&	0.056	&	0.33	&	0.045	\\
Cr	&	$	-9.72	^{+	0.25	}_{-	0.68	}	$	&		0.019	&	0.0070	&	0.0134	&	$	0.21	\pm	0.10	$	&	0.53	&	0.076	&	0.15	&	0.64	&	0.26	\\
Mn	&	$	-9.81	^{+	0.22	}_{-	0.46	}	$	&		0.015	&	0.0057	&	0.0084	&	$	0.17	\pm	0.08	$	&	0.53	&	0.065	&	0.13	&	0.57	&	0.19	\\
Fe	&	$	-7.51	^{+	0.19	}_{-	0.36	}	$	&		3.12	&	0.33	&	1.72	&	$	34.3	\pm	9.2	$	&	0.54	&	13.0	&	26.5	&	80.2	&	18.6	\\
\hline																																	
Li	&	$<$	-7.5							&	$<$	3.2	&		&		&	$<$	34.7				&	1.15	&	\ldots	&	\ldots	&	\ldots	&	1.5E-04	\\
N\tablenotemark{\ddag}	&	$<$	-7.3							&	$<$	5.0	&		&		&	$<$	55.0				&	0.86	&	\ldots	&	\ldots	&	\ldots	&	0.25	\\
Na	&	$<$	-7.2							&	$<$	6.3	&		&		&	$<$	69.2				&	0.81	&	\ldots	&	\ldots	&	\ldots	&	0.51	\\
Al	&	$<$	-7.7							&	$<$	2.0	&		&		&	$<$	22				&	0.79	&	0.56\tablenotemark{\dag}	&	1.13\tablenotemark{\dag}	&	$< $ 0.4	&	0.84	\\
V	&	$<$	-9.5							&	$<$	0.03	&		&		&	$<$	0.3				&	0.51	&	\ldots	&	\ldots	&	\ldots	&	5.4E-03	\\
Ni	&	$<$	-8.3							&	$<$	0.5	&		&		&	$<$	5.5				&	0.54	&	0.77\tablenotemark{\dag}	&	1.56\tablenotemark{\dag}	&	$< $ 10	&	1.10	\\
\hline						
\enddata						
\tablenotetext{\dag}{Assumed contribution (see text Section \ref{sec:oexcess})}
\tablenotetext{\ddag}{Nitrogen upper abundance limit from FUSE.  N is detected in an HST/COS spectrum (PI: B. G\"{a}nsicke) from which we derive an abundance of log(N/He) = -8.15 (see text Section \ref{sec:oexcess}).} 
\tablecomments{Similar to Table \ref{tab:2339_abund}, but for GD 378.  Upper limits are derived from non-detections of Li {\small I} 6708\,{\AA}, N {\small I} 1134\,{\AA}, Na {\small I} 5890\,{\AA}, Al {\small II} 3587\,{\AA}, V {\small II} 3125\,{\AA}, and Ni {\small I} 3515\,{\AA}.  The compositions by mass in the four columns on the right are the following (see also Section \ref{sec:accdiff}): ``steady state'' = accretion-diffusion equilibrium ; ``Rock only'' = the steady state phase with `excess' O removed from the O abundance and attributed to water ice; ``Decreasing Phase'' = the decreasing phase after $\sim$ 6-7 Be settling times; ``CI chondrites'' = meteoritic abundances from \citet{lodders2020}. 
}
\end{deluxetable*}

\begin{deluxetable*}{lrllrrllll}
\tablecaption{Beryllium Upper Limits and Major Element Abundances in Heavily Polluted WDs \label{tab:compare_abund}}
\tablehead{
\colhead{WD name}	&	\colhead{$T_{\rm eff}$}	&	 \colhead{log $g$}	&	\colhead{[H/He]} 	&		\colhead{[Be/He]} 	&		\colhead{[O/He]} 	&	\colhead{[Mg/He]} 	&	\colhead{[Si/He]} 		&	\colhead{[Fe/He]} 	&	\colhead{reference}	
}																						
\startdata																						
PG 1225$-$079	&	9940	&	7.97	&	-3.98	&	$<$	-12.0	&		\ldots	&	-7.43	&	-7.50		&	-7.52	&	this paper	\\
GD 362	&	10540	&	8.24	&	-1.14	&	$<$	-10.7	&	$<$	-5.14	&	-5.98	&	-5.84		&	-5.65	&	Z2007	\\
SDSSJ1242+5226	&	10710	&	7.93	&	-3.77	&	$<$	-11.0	&	$\sim$	-4.0	&	-5.68	&	-5.55		&	-6.11	&	this paper	\\
SDSSJ0738+1835	&	13950	&	8.40	&	-5.73	&	$<$	-10.0	&		-3.81	&	-4.68	&	-4.90		&	-4.98	&	D2012	\\
G200-39	&	14490	&	7.95	&	-4.2	&	$<$	-11.3	&		-6.62	&	-8.16	&	-8.03		&	-8.15	&	X2017	\\
Ton 345	&	18700	&	8.00	&	-5.1	&	$<$	-9.5	&		-4.58	&	-5.02	&	-4.91		&	-5.07	&	J2015	\\
\enddata
\tablecomments{Logarithmic abundances and upper limits by number.  Be abundance upper limits are all newly derived in this work.  Atmospheric parameters and abundances for O, Mg, Si, and Fe are from the papers listed in the reference column, except for the two that have been re-fit in this paper: PG 1225$-$079 (previously analyzed by \citet{klein2011} and \citet{xu2013gd362pg1225}), and SDSSJ1242+5226 (analyzed by \citet{raddi2015}; see also discussion in text Section \ref{sec:abund}).     References are: Z2007 \citep{zuckerman2007}; D2012 \citep{dufour2012}; X2017 \citep{xu2017};  J2015 \citep{jura2015}.  }
\end{deluxetable*}

\subsection{Accretion-Diffusion \label{sec:accdiff}}
Relating the detected atmospheric abundances to the parent body composition  depends on the interplay between accretion and diffusion in the system.   \citet{koester2009} describes a three-phase model of increasing, steady state, and decreasing abundances.  In the increasing phase settling is not yet playing a significant role, and the relative parent body abundances are directly the measured ones. For the steady state we need to take into account the different settling times ($\tau$(Z)) of the various elements using Equation 7 of \citet{koester2009} and $\tau$(Z) from Tables \ref{tab:2339_abund} and \ref{tab:gd378_abund}. This effect can modify the observed abundance ratios by up to factors of two.  We note that Figure 8 in \citet{heinonen2020} shows that improved physics models can lead to diffusion timescale ratios that, for $T_{\rm eff} \lesssim$10,000\,K, are substantially different from those currently available and used in the present paper. However, those differences remain small in the $T_{\rm eff} \sim$13,000-16,000\,K regime of the two Be-enriched WDs, so this uncertainty does not affect our conclusions in any way.

\citet{koester2009} has modeled the start of the decreasing phase by switching accretion off, at which point the heavy elements begin differentially settling in the WD atmosphere according to an exponential decay governed by their relative diffusion times.  We use this approach here, as it is sufficiently illustrative for the current paper, but we note that a somewhat more elaborate model includes an exponential decay of the accretion from the disk as described by \citet{jura2009apj}, and employed by \citet{doyle2020, doyle2021}.

The WDs studied here have settling times of the order 10$^5-10^6$ yrs, which is in the estimated range of disk lifetimes \citep{girven2012, verasheng2020}, so at the outset it is possible for us to interpret these systems to be in any one of the three phases.  A detected dust disk would narrow down the possibilities$-$to increasing phase or steady state$-$but as shown in Figures \ref{fig:j2339_sed} and \ref{fig:gd378_sed}, neither GALEXJ2339 nor GD 378 have a detected infrared excess.  Decreasing phase abundances can differ significantly from the original parent body composition as the element ratios undergo an exponential decay over time since the end of accretion. Since, aside from H, the lightest detected element, Be, has the longest dwell time compared to the other elements in the WD atmosphere, it is natural to ask, can a decreasing phase explain the large overabundance of Be?  

We calculated the number of e-folding settling times it would take to have a Be \% mass composition $-$ which began as chondritic $-$ evolve into what is observed today.  For both stars it would take about 7-8 Be settling times ($\sim$15 Fe settling times) to achieve the two orders of magnitude over-abundance of mass composition of Be seen in the WD atmospheres.  However, such a scenario also results in abundance patterns for the other elements that are extreme (see columns ``Decreasing Phase'' in Tables \ref{tab:2339_abund} and \ref{tab:gd378_abund}), requiring parent body compositions that are $\simeq$80\% Fe, with $\simeq$3-6\% O,  $\simeq$3\% Ca, $<$1\% in each of Mg, Si, \& Al, all with a chondritic proportion of Be. 
This bizarre makeup, together with the fact that for GALEXJ2339 after 7-8 settling times the mass of the polluting parent body would need to have started out more than three (probably closer to ten) times the mass of planet Earth, makes the decreasing phase a highly unlikely explanation for the large overabundance of Be in the two WDs.  
A thorough and quantitative treatment of the accretion-diffusion situation, and consideration of multiple accretion events, is presented in the companion to this paper by \citet{doyle2021}.

As mentioned above, the abundance ratios between the increasing phase and the steady state are quite similar, differing by at most a factor of two.   In the following analysis, for simplicity we consider only the steady state values, which is a conservative approach in this situation where we are dealing with the overabundance of lighter elements (longer settling times) compared to the other elements associated with the polluting planetary material.

\subsection{Oxygen Excesses \label{sec:oexcess}}
Besides the huge overabundance of Be, and large amounts of H, O is overabundant in both WDs as well. From Tables \ref{tab:2339_abund} and \ref{tab:gd378_abund} one finds that the number of O atoms are much greater than required to bond with the other major rock-forming elements (Mg, Si, and Fe), as detailed in the following paragraph. Two principle mechanisms can be responsible for excess O: (1) the system may be in a settling phase that makes O appear over-abundant; or (2) there was water in the accreted parent body.  We have already noted that a long-time declining phase is unlikely; it would take at least four Be settling times, $\sim$10 Myr and $\sim$4 Myr in GALEXJ2339 and GD 378 respectively, to account for their oxygen excesses.  By that point, the ratios of other elements become unlike any understandable abundance pattern.  We also pointed out that the large H abundance in these types of WDs implies the accretion of water, so next we evaluate that scenario.

  We use the $n(Z)/n(\rm O)$ abundance ratios to calculate oxygen budgets to get a measure of the partitioning of O in rocky material.  That is, as described in \citet[][Section 4.3]{klein2010}, we count up the number of O atoms that can be carried by the major and minor oxides in a rocky body: MgO, Al$_2$O$_3$, SiO$_2$, CaO, FeO, and NiO\footnote{Oxides from trace elements (Na$_2$O, P$_2$O$_5$, TiO$_2$, V$_2$O$_5$, Cr$_2$O$_3$ and MnO) can be included, but in practice their contributions to the oxygen budget are negligible.}. 
 
 Being among the top seven most abundant elements in bulk Earth, Al and Ni are expected to be non-negligible in a total parent body composition at the level of one to a few percent, but we only have upper limits on their observed abundances.  Therefore in calculating the \% mass compositions and oxygen budgets, we assume values for Al and Ni as associated with their partner elements, Ca and Fe, respectively. That is, for the increasing and steady state accretion phases we set the Al and Ni abundances to CI chondrite ratios (similar to the Sun and bulk Earth) at  Al/Ca = 0.94 and Ni/Fe = 0.058, by mass, since Al/Ca and Ni/Fe have similar behavior in rocky bodies based on their condensation temperatures and tendencies to be in a metallic form or not.
  
  For GALEXJ2339 and GD 378 we find that in the steady state, just 33\% and 28\%, respectively, of the detected O atoms could have been delivered in the form of rocky oxides. There is far more than enough hydrogen in the convection zones of these WDs to account for the excess O to be associated with H$_2$O ice.
After subtracting the excess O from its total abundance, except for Be the remaining overall mass composition patterns are remarkably similar to chondritic (see columns `Rock only no H$_2$O' and `CI Chondrites' in Tables \ref{tab:2339_abund} and \ref{tab:gd378_abund}). 
Considering relative proportions, the parent bodies polluting both WDs each contained roughly comparable amounts (by mass) of rock and water ice.
That, along with the observation of other volatile species in the FUSE spectrum of GD 378$-$namely C and S$-$leads us to conclude that the parent bodies which polluted these WDs were composed of material that originated beyond the ice-lines of their protoplanetary disks.

We note that an archival HST/COS spectrum of GD 378 (program \#12474) contains photospheric lines of nitrogen (B. G\"{a}nsicke, private communication), which is consistent with the results that we have established here. GD 378 is the second white dwarf known to display N from planetary accretion; G200-39 was the first \citep{xu2017}.  Referring to Table \ref{tab:gd378_abund} footnotes, the N abundance we derive from the COS spectrum is log(N/He)= -8.15, which translates to a steady state N mass composition of 0.5\% (or 1.0\% of the rocky material without H$_2$O). These percentages may be compared to 0.25\% in CI chondrites \citep{lodders2020}, 1.5\% in comet Halley's dust \citep{jessberger1988}, and $\simeq$2\% in the Kuiper Belt analog accreted by G200-39 \citep{xu2017}.

\begin{figure*}
\begin{center}
\includegraphics[width=170mm]{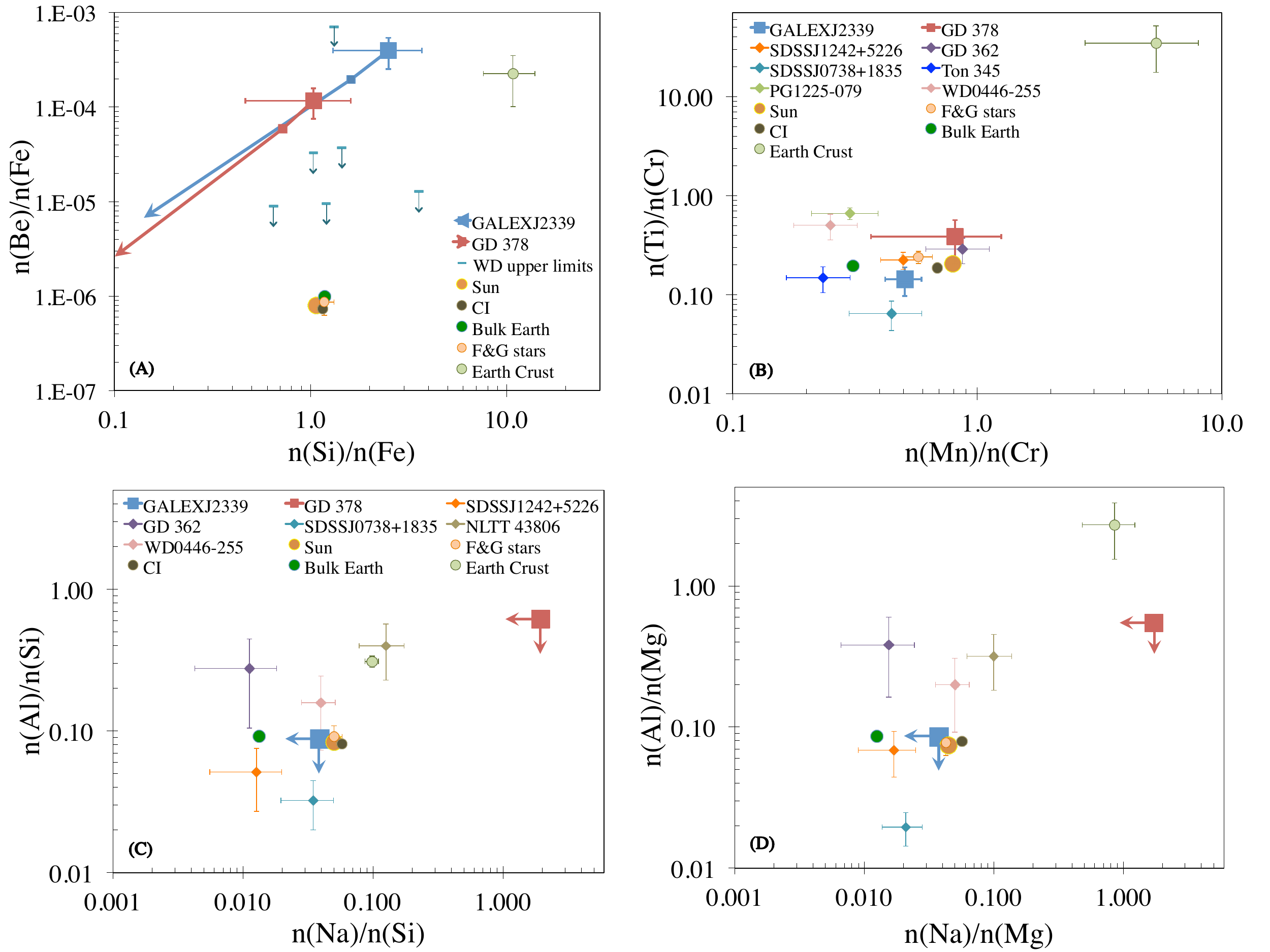}
\caption{Abundance ratios by number.  Large filled squares with error bars and/or upper limits are observed photospheric abundances in the two Be WDs, small squares in panel (A) mark the steady state values.  Following \citet{swan2019} we use backward time arrows to chart how the diffusion evolution would look if the systems are in a decreasing state.  The arrow head positions denote what the starting abundance ratios would have been if accretion ceased approximately 4 Be settling times ago (about 10 Myr for GALEXJ2339 and 4 Myr for GD 378).   In panels (B), (C), and (D) differential diffusion has almost no effect on the displayed ratios due to similar settling times of the plotted elements; these ratios are nearly independent of the accretion-diffusion states of the systems.  In panels (C) and (D) the arrows on GALEXJ2339 and GD 378 indicate upper abundance limits, and symbols in panel (D) are the same as those defined in (C). Filled circles are F- \& G-type stars (Be from \citet{boesgaard1976be}, all other elements from \citet{reddy2003}), solar and CI chondrites \citep{lodders2020}, bulk Earth \citep{allegre2001} and Earth's crust \citep{rudnickgao2003}.  WD upper limits for Be are from Table \ref{tab:compare_abund} and are uncorrected for settling.  Other WD abundances, uncorrected for settling, are from:  SDSSJ1242+5226 (this paper), GD 362 \citep{zuckerman2007}, SDSSJ0738+1835 \citep{dufour2012}, Ton 345 \citep{jura2015}, PG 1225$-$079 \citep{xu2013gd362pg1225}, WD 0446$-$255 \citep{swan2019}, NLTT 43806 \citep{zuckerman2011}. }
\label{fig:ratios}
\end{center}
\end{figure*}

\subsection{Comparison to Li Polluted WDs \label{sec:li}}
Two recent papers report the discovery of lithium in a handful of very cool (T$_{\rm eff}$ $<$ 5000\,K) WD atmospheres \citep{kaiser2020, hollands2021}.  The Li abundances appear modestly enhanced in all the stars.  \citet{kaiser2020} interpret the enhanced Li in these old WDs in the context of galactic chemical evolution, while \citet{hollands2021} assert that the observed Li in all four of their sample stars comes from differentiated planetary crusts.  Indeed, the elements Li and Be are enhanced in Earth's crust, but we also note that the overall composition of the crust is considerably different from bulk Earth, CI chondrites, and main sequence stars.  It is dominated by O, Si, Al, and Ca, enhanced in elements such as Li, Be, Na, Mn, and Ti, while depleted in elements such as Fe, Mg, and Cr.  
We have already pointed out in the preceding section, that the rocky portions of both parent bodies have similar overall abundance patterns as CI chondrites, but since we have measured suites of elements in both the Be WDs, we can evaluate some of the element ratios in more detail. See also \citet{doyle2021} for additional analysis of parent body compositions.

DB type WDs are generally too warm to allow the detection of easily ionizable elements such as Li and Na, but nonetheless we derive their upper limits (Tables \ref{tab:2339_abund} and \ref{tab:gd378_abund}).  A combination of the heavy pollution and cooler temperature of GALEXJ2339 yields an upper limit of Na that is close to chondritic, while in GD 378 the Na limit is much less constrained.  For lithium, the upper limits are near or much higher than the overabundances observed for Be in the two stars: in GALEXJ2339 the steady state Li/Mg upper limit is $\simeq$300 times the chondritic ratio, while in the case of GD 378 it is more than 5000 times chondritic.  Thus it is possible that Li may also be extremely overabundant in either or both of GALEXJ2339 and GD 378, but we are simply not able to measure it with current observations. 

Referring to Figure \ref{fig:ratios}, panel (A) shows how the Be/Fe ratios may be compared to crust, but the Si/Fe ratios do not agree. In panel (C) the unique similarity with Earth crust ratios is apparent for NLTT 43806, whose Al-rich abundance pattern has been interpreted as originating from a parent body containing a significant amount of crust \citep{zuckerman2011}.  The other WDs do not display such a pattern, and combined with the comparisons shown in panels (B) and (D), it is clear that the abundances of the two Be WDs are not at all similar to planetary crust material.   Rather, apart from the Be overabundance, the element ratios and rocky mass compositions of these two WD are, by and large, consistent with chondritic.

\section{Discussion \& Conclusions \label{sec:discussion}}

Ann Merchant Boesgaard has spearheaded studies of light element abundances in stars in the solar vicinity.   Summaries can be found in \citet{boesgaard1976light} and \citet{boesgaard2020}.  The primary production mechanism for boron and beryllium is generally thought to be spallation by cosmic rays on elements such as oxygen, carbon, and nitrogen in the interstellar medium.  For lithium, production by spallation is important, but not necessarily the dominant production process. Abundances of lithium, beryllium, and boron in young main sequence stars are typically comparable to or greater than abundances in older main sequence stars.  Li abundances are enhanced in a few post-main sequence stars, but this has never been found to be the case for beryllium; Be abundances are always reduced by stellar evolution.

Table 3 in \citet{boesgaard1976be} gives mean Be abundances by number relative to hydrogen in main sequence G- and F-type stars (including the Sun) equal to 1.3 x 10$^{-11}$.  Boesgaard remarks that, within the errors, the meteoritic value agrees with the solar value and that 1.3 x 10$^{-11}$ is the cosmic Be abundance.  Table 2 in \citet{boesgaard1976be} contains 38 G- and F-type stars, none of which have a Be abundance more than a factor of two above solar.  A mostly independent set of stars is plotted in Figure 5 of \citet{boesgaard2004} where the most Be-rich stars have Be abundances that are again within a factor of two of solar and agree with the meteoritic value.

Much larger downward deviations from the mean are found in some main-sequence F-type stars with temperature in the range 6400 to 6800 K; by Hyades age Be can be depleted by up to a factor of six \citep{boesgaard2020}.  The progenitors of many white dwarfs are such F-type stars.

Given the above, we conclude that the excess of Be in GD 378 and GALEXJ2339 is a signature of an environment where O and/or C, N and protons were subjected to MeV collisions, either direct (accelerated protons) or reverse (accelerated O and/or C, N), resulting in unusually efficient production of Be by spallation. The high flux was presumably the result of proximity to the source of energetic particles and a stopping distance comparable to the scale of the targets. One possibility is that the star and planetesimal formation may have occurred in an environment containing a strong source of high energy radiation.  If the high-energy source were interior to the protoplanetary disk, i.e.~the star itself, then the irradiated gas would likely be sufficiently close to the star that planetesimals formed out of the gas would be relatively dry (unlike the ice-rich bodies found here), and  could not survive the evolution of the star unless they migrated to larger semi-major axes. Also, if irradiated gas with a high Be abundance located close to a star were to accrete onto the outer layers of that star, then one might expect to see at least some young stars with supersolar Be abundances. But no such stars have been discovered in the Pleiades and $\alpha$Per clusters \citep[e.g.][]{boesgaard2003}. 

The foregoing assumes a quiescent protoplanetary disk, but different scenarios could occur in more dynamic disks, such as those with magnetically driven outflows.  Magnetocentrifugal winds can potentially transport thermally-processed irradiated material from inner to outer regions of a protoplanetary disk \citep[e.g.~][]{shu1997, giacalone2019}, where it may become incorporated into planetesimals formed at those locations.  On the other hand,   if the source was external to the progenitor (for example, a nearby Wolf Rayet star) then it could irradiate the outer gaseous portions of a protoplanetary disk before rocky planetesimals were fully formed.  In either of the last two scenarios, such planetesimals then could have survived until the star evolved into a white dwarf, and would potentially contain significant amounts of water ice, as found in the objects studied here.

An alternative model to explain the high Be abundance is presented in the accompanying paper by \citet{doyle2021}.  Whatever the case, if the measured high Be abundance is due to a spallation process of any sort, then one also anticipates enhanced lithium and boron abundances.

More generally, this remarkable detection of Be suggests that, especially with the next generation of large telescopes, additional elements may be added to Table \ref{tab:history}, providing new insights into processes associated with planetary formation and/or evolution.  For example, an observation of barium in a polluted WD could inform about the presence (or lack of) plate tectonics in an extrasolar planetary body as predicted by \citet{jura2014}. \\

\acknowledgements

ACKNOWLEDGEMENTS

This paper is dedicated to the memory of Michael Jura, who was a pioneer and leader in the study of polluted white dwarfs, especially in the context of measuring and understanding the compositions of extrasolar minor planets.  Fittingly, asteroid 6406 Mikejura (\url{https://ssd.jpl.nasa.gov/sbdb.cgi?sstr=6406;orb=1}) was named after him. 

We thank the anonymous referee for a helpful report which improved the manuscript.
This paper has benefitted from conversations with Detlev Koester, Boris G\"{a}nsicke, Jay Farihi, Geoff Marcy, and Ted Johnson. We are grateful to Boris G\"{a}nsicke for kindly giving us permission to discuss the unpublished detection of photospheric nitrogen in GD 378's COS spectrum.  We also thank Diego Romani for assisting with the optical spectrum of GD 378 and Lou Baya Ould Rouis for help with observing planning. 
 
B.K. was supported by the A.P.S.~M. Hildred Blewett Fellowship. 
A.D. and B.K.~acknowledge support from the NASA Exoplanets program award number 80NSSC20K0270 (EDY).  Partial support was also provided by other NASA grants to UCLA.  S.B.~acknowledges support from the Laboratory Directed Research and Development program of Los Alamos National Laboratory under project number 20190624PRD2. C.M. and B.Z.~acknowledge support from NSF grants SPG-1826583 and SPG-1826550.

Some of the data presented herein were obtained at the W.M. Keck Observatory, which is operated as a scientific partnership among the California Institute of Technology, the University of California and the National Aeronautics and Space Administration. The Observatory was made possible by the generous financial support of the W.M. Keck Foundation.  We recognize and acknowledge the very significant cultural role and reverence that the summit of Mauna Kea has always had within the indigenous Hawaiian community.

This work has made use of data from the European Space Agency (ESA) mission
{\it Gaia} (\url{https://www.cosmos.esa.int/gaia}), processed by the {\it Gaia}
Data Processing and Analysis Consortium (DPAC,
\url{https://www.cosmos.esa.int/web/gaia/dpac/consortium}). Funding for the DPAC
has been provided by national institutions, in particular the institutions
participating in the {\it Gaia} Multilateral Agreement.

This publication makes use of data products from the Two Micron All Sky Survey, which is a joint project of the University of Massachusetts and the Infrared Processing and Analysis Center/California Institute of Technology, funded by the National Aeronautics and Space Administration and the National Science Foundation.
The following atomic spectral line databases were consulted: Vienna Atomic Line Database (VALD), Kurucz (1995, R.L. Kurucz and B. Bell, CD-ROM No. 23, Cambridge, Mass.: Smithsonian Astrophysical Observatory), NIST Standard Reference Database 78, and \citet{vanhoof2018}.

\appendix

\section{Abundance Fitting}
The abundances are extracted as follows.  For a range of effective temperatures, surface gravities, and [H/He] abundances, we varied the abundances of all the other detected elements in steps of 0.5 dex.  We use the Vienna Atomic Line Database (VALD) for the atomic data of each line\footnote{Atomic line data for the well-studied Be lines are almost identical in other atomic databases, Kurucz, NIST, and \citet{vanhoof2018}.}.  This provides us a grid of model atmospheres and synthetic spectra which we then interpolate to fit the final abundances. We follow an approach similar to that described in \cite{dufour2012} and divide the observed spectra into 5-10\,{\AA} segments that are centered around the  spectral lines that we want to fit. For each of those segments, we use a $\chi^2$ minimization algorithm to find the abundance that yields the best fit to the line(s) present in the segment. Only one element at a time is fitted in each segment and most elements are fitted in more than one segment. Two contributions dominate the uncertainty on the absolute abundances: the uncertainty on $T_{\rm eff}$ and the spread between abundances derived for different segments.  The first contribution is obtained by performing the fitting procedure at $T_{\rm eff} \pm \sigma({T_{\rm eff}})$ and the second contribution is estimated by taking the standard deviation of the mean from the different abundance measurements. If an element only appears in one or two segments, then the ``spread'' error is estimated by inspection of model fits at varied higher and lower abundances. 

It has been shown that the element-to-element ratios are much less sensitive to variations in $T_{\rm eff}$/\,log $g$ than are the absolute abundances  \citep{klein2010, klein2011}. Nonetheless, we want to estimate both the absolute and relative abundance dependence on $T_{\rm eff}$/\,log $g$ \footnote{Note that $T_{\rm eff}$ and \,log $g$ are themselves linked through the flux-solid angle formula used in photometric-parallax fitting.  That is, with the distance fixed, a hotter model requires a smaller $R_{\rm WD}$ (larger $M_{\rm WD}$, i.e.~larger \,log $g$) to fit the photometry \citep[see e.g.~Equation 1 of][]{coutu2019}.}.   But first, some comments on dealing with abundances in log-space versus number-space are in order.  When discussing absolute abundances of elements that can be anywhere between 1 to 11 orders of magnitude less abundant than the dominant atmospheric element (H or He), it is certainly convenient to use logs: [Z/H(e)] = log$_{10}$[$n(Z)/n(\rm H(e))$].  However, if one has to convert a log uncertainty to number-space, the resulting uncertainties will be asymmetric about the nominal number abundance.  This can cause difficulties in error propagation if one wishes to calculate a quantity in number-space, particularly those involving element-to-element ratios such as an oxygen budget or fugacity \citep{doyle2019, doyle2020}.  In our model fitting, we measure the element abundances in number-space, $n(Z)/n(\rm H(e))$, and we also calculate the spread uncertainties in number-space as described in the preceding paragraph.  Thus, here we choose to report abundances and their symmetric uncertainties in number-space (which translates to asymmetric uncertainties in log-space).

To separate the model-dependent (i.e.~$T_{\rm eff}$/\,log $g$) uncertainty contributions from those of spectral measurements, in Tables  \ref{tab:2339_abund} and \ref{tab:gd378_abund}, we give abundances with the associated contributions from the spread error $\sigma_{\rm spread}$ and the error from varying the temperature, $\sigma_{T_{\rm eff}}$, listed separately.  This way, the uncertainties on relative element abundances can be computed by propagating the uncertainties on the individual element abundances while taking care to remove the correlated portion of the uncertainty related to $T_{\rm eff}$, according to Equation \ref{eq:sigma_ZtoZ}.

Explicitly, for general $Z_i$ and $Z_j$, the element-to-element abundance ratios, $n(Z_i)/n(Z_j)$, are obtained directly from the $n(Z_i)/n({\rm He})$. Continuing to work in number-space but dropping the ``n()'' for simplicity, the total uncertainty on the absolute abundance $Z_i/{\rm He}$ is just a propagation of the independent uncertainty contributions, $\sigma_{\rm spread}$ and $\sigma_{T_{\rm eff}}$:

\begin{equation}
    \sigma_{\rm tot}\left(\frac{Z_i}{\rm He}\right)  = \left( \sigma_{\rm spread}^2\left(\frac{Z_i}{\rm He}\right) + 
     \sigma_{T_{\rm eff}}^2\left(\frac{Z_i}{\rm He}\right)  \right)^{1/2},
  \label{eq:sigma_ZtoHe}
\end{equation}
and the uncertainty in the ratio $Z_i/Z_j$ may be calculated as:

\begin{equation}
    \frac{\sigma(\frac{Z_i}{Z_j})}{(\frac{Z_i}{Z_j})} = \left( \left(\frac{\sigma_{\rm spread}(\frac{Z_i}{\rm He})}{(\frac{Z_i}{\rm He})}\right)^2 + 
     \left(\frac{\sigma_{\rm spread}(\frac{Z_j}{\rm He})}{(\frac{Z_j}{\rm He})}\right)^2  +   
\left(\frac{\sigma_{T_{\rm eff}}(\frac{Z_i}{\rm He})}{(\frac{Z_i}{\rm He})} - \frac{\sigma_{T_{\rm eff}}(\frac{Z_j}{\rm He})}{(\frac{Z_j}{\rm He})} \right)^2  \right)^{1/2},
  \label{eq:sigma_ZtoZ}
\end{equation}

\noindent where the third term on the right-hand-side accounts for the fact that the set of element abundances predominantly move together, up and down, with variations in $T_{\rm eff}$/\,log $g$.  

This kind of treatment assumes the dominant abundance uncertainty comes from line measurement and modeling variations, but we are aware that there are systematic uncertainties that can be contributing to the measurements in a non-statistical way.  For example, the observation of self-reversed core inversions in the He {\small I} 5876 profile of DB WDs suggests that there may be some problems with model temperature calibrations, possibly because of missing physics such as 3D effects, convective overshoot, and NLTE effects \citep{klein2020}.   Also, discrepancies between UV and optical abundances noted for various stars \citep[e.g.][]{gaensicke2012cos, xudusty2019} may be due to uncertain atomic data \citep{vennes2011}.

\section{Model Checks From Abundances}

In GD 378, three elements (O, Si, Fe) are detected in both FUSE (UV) and HIRES (optical) data, so we began by analyzing those spectra separately to check for possible UV vs.~optical abundance discrepancies.  Referring to Table \ref{tab:uv-opt}, the UV and optical abundances derived for these three elements agree within the uncertainties. Given the UV-optical consistency for these three major elements, we proceeded with our abundance analysis on the combined HIRES + FUSE spectrum, generating a single model that incorporates the fits to all observed lines.   

\begin{table}[h!]
\caption{UV-Optical abundance comparison for GD 378 \label{tab:uv-opt}}
\begin{center}
\begin{tabular}{lcccc}
\hline 
\hline
Z	&	FUSE	&  HIRES	\\	
	&	log[n($Z$)/n(He)]	  & 	log[n($Z$)/n(He)]  &  \\
\hline
O	&	$	-6.12	^{+	0.20	}_{-	0.36	}	$	&	$	-5.94	^{+	0.19	}_{-	0.35	}	$	&	\\
Si	&	$	-7.43	^{+	0.14	}_{-	0.21	}	$	&	$	-7.58	^{+	0.04	}_{-	0.05	}	$	&	\\
Fe	&	$	-7.43	^{+	0.19	}_{-	0.36	}	$	&	$	-7.60	^{+	0.20	}_{-	0.37	}	$	&	\\
\hline
\end{tabular}
\end{center}
\end{table}

We also checked the ionization balance, i.e.~the degree of agreement in abundances derived from different ionization states of the same element.  For GD 378, the total Mg abundance, as derived separately from lines of Mg {\small I} and Mg {\small II}, only differ by 27\% at the nominal model temperature,  which is similar to what we find with the cooler model  (23\%)  and somewhat worse (45\%) from the hotter model.  Likewise, the discrepancy between the total Fe abundance from Fe {\small II} and Fe {\small III} is only 14\%, 11\%, and 10\% from the cool, nominal, and hot models, respectively.  Si {\small II} and Si {\small III} have larger differences of 98\%, 83\% and 64\% from the cool, nominal, and hot models, respectively.  The ionization balance of Mg favors the nominal and cooler models, while Fe and Si favor the nominal and hotter models.  In GALEXJ2339 total abundances from Mg {\small I} and Mg {\small II} differ by 50\%, 25\% and 30\% in cool, nominal, and hot models, respectively, while the agreement is excellent for Fe {\small I} and Fe {\small II} at 5\%, 1\%, and 28\%. Both elements have the best accord with the nominal $T_{\rm eff}$ model.   Thus, we find the ionization balance supports our best fit (nominal) $T_{\rm eff}$ for each of the two stars.


\bibliographystyle{aasjournal}
\bibliography{BKrefs}

\end{document}